\documentclass[%
 reprint,
groupedaddress,
 amsmath,amssymb,
 aps,
 pra,
floatfix,
]{revtex4-2}

\usepackage{graphicx}
\usepackage{subcaption}
\usepackage{dcolumn}
\usepackage{bm}
\usepackage{color}
\usepackage[T1]{fontenc}
\usepackage[utf8]{inputenc}
\usepackage{lmodern}
\usepackage{textcomp}
\usepackage{comment}
\usepackage{grffile}
\usepackage{algorithm}
\usepackage{algpseudocode}
\usepackage{physics}
\usepackage{float}
\usepackage{natbib}
\usepackage{silence}
\usepackage{ulem}
\newcommand{\hk}[1]{\textcolor[rgb]{0.0, 0, 0.0}{#1}}

\begin{document}

\preprint{APS/123-QED}

\title{Proposal for realizing quantum-spin systems on a two-dimensional square lattice with Dzyaloshinskii-Moriya interaction by Floquet engineering using Rydberg atoms}

\author{Hiroki Kuji}
    \email{1225702@ed.tus.ac.jp}
\author{Masaya Kunimi}
    \email{kunimi@rs.tus.ac.jp}
\author{Tetsuro Nikuni}
    \email{nikuni@rs.tus.ac.jp}
\affiliation{Department of Physics, Tokyo University of Science,1-3 Kagurazaka, Shinjuku, Tokyo 162-8601,  Japan}




\date{\today}

\begin{abstract}
We theoretically propose a method for implementing a Hamiltonian incorporating Heisenberg and Dzyaloshinskii-Moriya (DM) interactions within Rydberg atoms arranged in a two-dimensional square lattice, utilizing Floquet engineering. In our scheme, we use both global and local operations of the spins. The global operations can be realized by applying microwave pulses, and the local operations can be realized by locally addressing atoms with off-resonant lasers, which induce the site-dependent ac-Stark shift. Since our engineered Hamiltonian contains bond-dependent DM interactions, we expect the emergence of quantum skyrmions in the ground state.
\end{abstract}

\maketitle
\section{Introduction}\label{sec:Introduction}

Quantum simulation~\cite{lloyd1996universal,feynman2018simulating}, which involves the experimental emulation of specific physical systems, is a promising platform for addressing problems that are intractable with classical computers~\cite{abrams1999quantum}. This approach has significantly contributed to the exploration of various quantum many-body phenomena using known platforms such as ultracold gases~\cite{schafer2020tools}, trapped ions~\cite{monroe2021programmable,blatt2012quantum}, molecules~\cite{yan2013observation,zhou2011long}, and superconducting qubits~\cite{altman2021quantum,kjaergaard2020superconducting,wendin2017quantum}. Recently, a platform based on Rydberg atoms has attracted much attention due to its long coherence time, scalability, and high controllability via optical tweezers. Quantum simulations employing Rydberg atoms~\cite{browaeys2020many,weimer2010rydberg} have been utilized to investigate various quantum many-body phenomena, such as quantum phases~\cite{zhang2017observation,chen2023continuous,scholl2021quantum,ebadi2021quantum}, nonequilibrium dynamics~\cite{bernien2017probing}, and quantum thermalization~\cite{kim2018detailed,gross2017quantum}.

The Rydberg atom-based quantum simulator shows great potential for simulating quantum spin systems. However, it is limited to Hamiltonians that can be implemented. To date, one can experimentally realize quantum spin models as the Ising~\cite{labuhn2016tunable,zeiher2017coherent,bernien2017probing,de2018accurate,lienhard2018observing,guardado2018probing,keesling2019quantum,scholl2021quantum,ebadi2021quantum,semeghini2021probing,bluvstein2021controlling,PhysRevLett.131.123201}, {\it XY}~\cite{orioli2018relaxation,de2019observation,chen2023continuous}, {\it XXZ}~\cite{signoles2021glassy,scholl2022microwave,franz2022absence,PhysRevResearch.6.033131} and {\it XYZ}~\cite{geier2021floquet,PhysRevLett.130.243001} models. There are several theoretical proposals for realizing a Hamiltonian with mono-axial Dzyaloshinskii-Moriya (DM) interactions~\cite{PhysRevA.108.053318,kunimi2024,PhysRevA.108.023305,kuznetsova2023engineering} and Kitaev type interactions~\cite{PhysRevX.13.031008, PhysRevA.108.053318, chen2023realization}. In this paper, we focus on the DM interaction~\cite{dzyaloshinsky1958thermodynamic,moriya1960anisotropic}, which is defined by the outer product between two spins. This interaction is fundamental to various chiral magnetic structures, such as chiral solitons~\cite{togawa2012chiral,kishine2005synthesis} in one-dimensional systems and skyrmions~\cite{SKYRME1962556,bogdanov1989thermodynamically,mühlbauer2009skyrmion,tokura2020magnetic} in higher dimensions. 
In particular, the full quantum calculation of the quantum skyrmions is a challenging task because the DM interaction causes the negative sign problem. Although the tensor network approach has been used for quantum skyrmions, it suffers from numerical errors in large system sizes~\cite{PhysRevResearch.4.043113}. Quantum simulation using Rydberg atoms is an alternative approach for studying quantum skyrmions.

Floquet engineering is a powerful method for controlling Hamiltonian~\cite{PhysRevX.4.031027,bukov2015universal} and is used to engineer new effective Hamiltonians~\cite{scholl2022microwave,geier2021floquet, PhysRevA.108.053318, köylüoğlu2024floquetengineeringinteractionsentanglement}. Especially, previous studies have experimentally demonstrated the realization of the {\it XYZ} Hamiltonian using global time-periodic operations~\cite{scholl2022microwave,geier2021floquet}. A theoretical proposal has also been made to implement a mono-axial DM interaction~\cite{PhysRevA.108.053318} using local operations~\cite{de2017optical,PhysRevLett.132.263601}. In this previous work, Nishad {\it et al}. proposed a method to engineer the Hamiltonian with the monoaxial DM interactions in a one-dimensional chain via Floquet engineering~\cite{PhysRevA.108.053318}. However, the Hamiltonian with more complicated interaction, such as bond-dependent DM interaction, in higher dimensions has yet to be realized in the Rydberg atom quantum simulator.

In this paper, we propose a method to implement the Hamiltonian with Heisenberg and DM interactions in a Rydberg atom system arranged in a two-dimensional square lattice using the Floquet engineering approach. Our starting point is the {\it XY} Hamiltonian, which can be realized in the Rydberg atom systems~\cite{walker2005zeros,barredo2015coherent,ravets2014coherent}. By applying both time-periodic global and local pulses to the system, we can obtain the effective Hamiltonian that has the Heisenberg and bond-dependent DM interactions.  We verify that our effective Hamiltonian can correctly reproduce the stroboscopic dynamics of the original Hamiltonian by numerical calculations. Our proposed experimental scheme is feasible with current state-of-the-art experimental techniques.

This paper is organized as follows: In Sec.~\ref{sec:model}, we describe our model and theoretical methodology employed in this study. In Sec.~\ref{sec:results}, we explain specific pulse sequences for implementing the Hamiltonian with the Heisenberg and DM interactions. In Sec.~\ref{sec:numerical analysis}, we present numerical results that validate the proposed sequence for both the ideal and realistic cases. In Sec.~\ref{sec:summary}, we summarize our results. In the Appendixes, we discuss the details of the calculations, the long-range interactions, and the symmetry properties of the effective Hamiltonian.
\section{Model and Method}\label{sec:model}
\begin{figure*}[t]
    \centering
    \includegraphics[width=\textwidth]{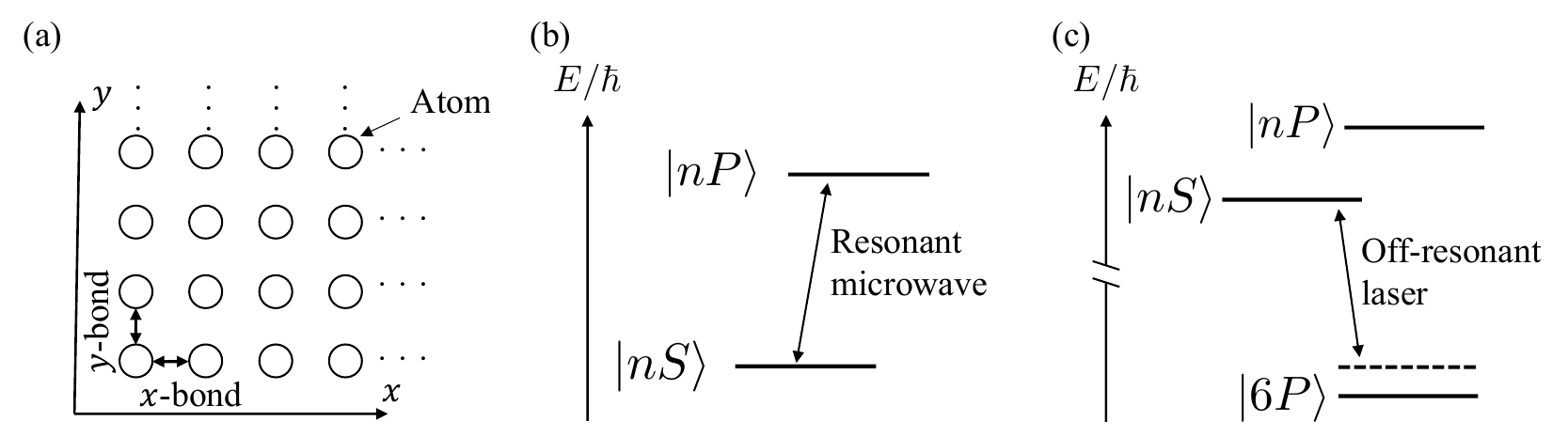}
    \caption{(a) Schematic of the system in this study. (b) Schematic for global operations using microwave pulses. We can rotate all spins simultaneously by applying resonant microwave pulses, which couple $|nS\rangle$ and $|nP\rangle$ states. This operation corresponds to the uniform spin rotation around the $x$ or $y$ axis. (c) Schematic for the local operation. By applying the off-resonant laser pulse to the individual atom, we can induce the site-dependent ac-Stark shift. This operation corresponds to the site-dependent spin rotation around the $z$ axis.}
    \label{fig:schematic diagram}
\end{figure*}
In this study, we consider a system consisting of Rydberg atoms arranged in a two-dimensional square lattice with lattice spacing $a$. See Fig.~\ref{fig:schematic diagram}(a). The position of the $i$th site is defined by $\bm{R}_i\equiv a(n_i\bm{e}_x+m_i\bm{e}_y)$, where $n_i$ and $m_i$ are integers, and $\bm{e}_{x,y}$ is the unit vector of each direction. We consider the spin-1/2 quantum spin systems. The two Rydberg states are mapped to the spin states $\ket{nS}=\ket{\uparrow}$ and $\ket{nP}=\ket{\downarrow}$~\cite{orioli2018relaxation,de2019observation,geier2021floquet,scholl2021quantum,chen2023continuous}. The states $\ket{\uparrow_j}$ and $\ket{\downarrow_j}$ denote the eigenstates of $\hat{S}^z_{\bm{R}_j}$ with eigenvalues $+1/2$ and $-1/2$, respectively, where $\hat{S}^{\alpha}_{\bm{R}_j}\;(\alpha=x,y,z)$ is the spin-1/2 operator at the $i$th site. Because these two Rydberg states have opposite parity, the dipole-dipole interaction has nonzero matrix elements in the first-order perturbation for sufficiently large lattice spacing. In this setup, the Hamiltonian is given by the dipolar {\it XY} model~\cite{de2017optical,orioli2018relaxation,de2019observation,browaeys2020many,chen2023continuous,chen2023spectroscopy}:
\begin{align}
    \hat{H}_{0}=\frac{1}{2}\sum_{i,j,i\neq j}J_{ij}(\hat{S}^x_{\bm{R}_{i}}\hat{S}^x_{\bm{R}_{j}}+\hat{S}^y_{\bm{R}_{i}}\hat{S}^y_{\bm{R}_{j}}),
    \label{eq1}
\end{align}
where $J_{ij}\equiv 2C_3/|\bm{R}_i-\bm{R}_j|^3$ is the strength of the dipole-dipole interaction between the $i$th and $j$th sites, and $C_3$ is the interaction constant. In this paper, we use open boundary conditions, and assume that the dipole-dipole interaction is isotropic, which can be realized by setting the quantization axis perpendicular to the {\it xy} plane~\cite{weimer2010rydberg,browaeys2020many}.

We use Floquet engineering ~\cite{PhysRevX.4.031027,bukov2015universal} to realize the desired interaction Hamiltonian. In this scheme, we apply a time-periodic external field and focus on a timescale longer than the period of the external field. We consider the {\it XY} model Hamiltonian with an external field:
\begin{align}
    \hat{H}(t)=\hat{H}_{0}+\hat{H}_{\rm{drive}}(t),\label{eq:total_Hamiltonian}
\end{align}
where $\hat{H}_{\rm{drive}}(t)$ represents the time-periodic external field. In Floquet engineering, the time-evolution operator is defined as

\begin{align}
\hat{U}(t)&\equiv \mathcal{T}\exp\left[-\frac{i}{\hbar}\int^t_0dt'\hat{\tilde{H}}(t')\right],\label{eq:definition_of_time_evolution_operator}\\
\hat{\tilde{H}}(t)&\equiv \hat{U}_{\rm drive}^{\dagger}(t)\hat{H}_0\hat{U}_{\rm drive}(t),\label{eq:definition_of_H_tilde_t}\\
\hat{U}_{\rm drive}(t)&\equiv \mathcal{T}\exp\left[-\frac{i}{\hbar}\int^t_0dt'\hat{H}_{\rm drive}(t')\right],\label{eq:definition_of_U_drive}
\end{align}
where $\mathcal{T}$ denotes the time-ordered product, and $\hbar$ is the reduced Planck constant. The operator $\hat{\tilde{H}}(t)$ represents the Hamiltonian in the rotating frame. From the time-evolution operator (\ref{eq:definition_of_time_evolution_operator}), we can define the Floquet Hamiltonian $\hat{H}_{\rm F}$ as
\begin{align}
    \hat{U}(T)=e^{-i\hat{H}_{\rm{F}}T/\hbar},\label{eq:definition_of_Floquet_Hamiltonian}
\end{align}
where $T$ is the period of the external field.

The Floquet Hamiltonian reproduces the same time evolution at $t=nT$ as the Hamiltonian $\hat{\tilde{H}}(t)$, where $n$ is an integer. In general, it is difficult to obtain the exact expression for the Floquet Hamiltonian. To do this, we use the Floquet-Magnus expansion~\cite{bukov2015universal,kuwahara2016floquet}. The leading term of the Floquet Hamiltonian is given by
\begin{align}
     \hat{H}_{\mathrm{F}}^{(0)}=\frac{1}{T}\int_{0}^{T}dt\hat{\tilde{H}}(t).
\end{align}
Since this expansion is a short-time expansion, the timescale of the interaction is much longer than the period $T$, which implies ${\rm max}_{i,j}(|J_{ij}|) T/\hbar \ll 2\pi$.

By choosing the drive Hamiltonian $\hat{H}_{\rm drive}(t)$ appropriately, we can obtain a desired effective Hamiltonian. To do this, we use two types of external fields: global operation, which uses microwave pulses to manipulate all atoms simultaneously, and local operation, which uses laser pulses to manipulate each atom individually~\cite{PhysRevA.108.053318}. See Figs.~\ref{fig:schematic diagram}(b) and \ref{fig:schematic diagram}(c). To implement such pulses, we consider the following drive Hamiltonian: 
\begin{align}
    \notag\hat{H}_{\rm{drive}}(t)=\hat{H}_{\rm{G}}(t)+\hat{H}_{\rm{L}}(t)&\\
    \notag=\hbar\Omega(t)\sum_i[\cos\phi&(t)\hat{S}^x_{\bm{R}_i}+\sin\phi(t)\hat{S}^y_{\bm{R}_i}]\\
    &+\hbar\sum_i\Delta_i(t)\hat{S}^z_{\bm{R}_i},
\end{align}
where $\hat{H}_{\rm{G}}(t)$ is the global pulse term and $\hat{H}_{\rm{L}}(t)$ is the local pulse term, which are defined by
\begin{align}
    \hat{H}_{\rm{G}}(t)&\equiv \hbar\Omega(t)\sum_i[\cos\phi(t)\hat{S}^x_{\bm{R}_i}+\sin\phi(t)\hat{S}^y_{\bm{R}_i}],\\
    \hat{H}_{\rm{L}}(t)&\equiv \hbar\sum_i\Delta_i(t)\hat{S}^z_{\bm{R}_i}.
\end{align}
$\Omega(t)$ is the time-dependent Rabi frequency of the drive, $\phi(t)$ is the phase, and $\Delta_i(t)$ is the space- and time-dependent ac-Stark shift. In this paper, we approximate the time dependence of the Rabi frequency $\Omega(t)$ and ac-Stark shift $\Delta(t)$ by delta functions, assuming that the pulses are applied to the system instantaneously. We denote the time interval between pulses by $\tau_i$. We apply a two-step rotating-frame transformation with respect to the pulses for both global and local operations~\cite{PhysRevA.108.053318}. Under the above settings, we obtain the effective Hamiltonian:
\begin{align}
    \hat{H}^{(0)}_{\rm{F}}=\frac{1}{T}\sum_i \hat{H}^{\rm{rot}}_i \tau_i,
    \label{eq:exprsssion_of_effective_Hamiltonian}
\end{align}
where $\hat{H}_i^{\rm rot}$ is the Hamiltonian in the rotating frame during the time interval $\tau_i$. We choose the pulse sequence so that the effective Hamiltonian (\ref{eq:exprsssion_of_effective_Hamiltonian}) becomes the desired Hamiltonian. Finally, the obtained pulse sequence in the rotating frame is transformed to derive the pulse sequence in the laboratory frame. 

In this work, we focus on the {\it XYZ} and DM interaction terms, which are defined by
\begin{align}
\hat{H}_{XYZ}&\equiv \frac{1}{2}\sum_{\mu=x,y,z}\sum_{i,j,i\not=j}J_{ij}^{\mu}\hat{S}^{\mu}_{\bm{R}_i}\hat{S}^{\mu}_{\bm{R}_j},\label{eq:definition_of_XYZ_term}\\
\hat{H}_{\rm DM}&\equiv \frac{1}{2}\sum_{i,j,i\not=j}\bm{D}_{ij}\cdot(\hat{\bm{S}}_{\bm{R}_i}\times\hat{\bm{S}}_{\bm{R}_j}  ),\label{eq:definition_of_DM_interaction_term}
\end{align}
where $J_{ij}^{\mu}$ is the interaction strength of each spin component, and $\bm{D}_{ij}=-\bm{D}_{ji}$ is the DM vector. The {\it XYZ} Hamiltonian can be realized experimentally by applying the time-periodic global microwave pulses~\cite{geier2021floquet,scholl2022microwave}. The method to create a monoaxial DM interaction ($\bm{D}_{ij}\propto\bm{e}_z$) in the Rydberg atoms has been theoretically proposed~\cite{PhysRevA.108.053318}. This method is based on applying the local spin rotation around the $z$ axis (see Appendix~\ref{sec:appendix_rotation} for details). Before presenting our results, we review the previous studies for the readers' convenience.

\begin{figure}[H]
    \centering
    \includegraphics[width=8cm]{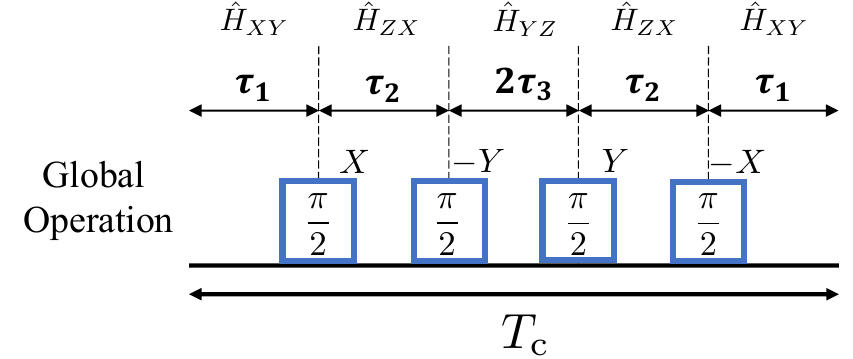}\\
    \caption{Pulse sequence for creating the $XYZ$ model Hamiltonians. Here, we present a sequence with one cycle period $T_{\rm c}$. Global operation means applying the pulse to all trapped atoms. The angle $\pi/2$ denotes the rotation angle in the laboratory frame. The symbols $X,Y,-X,$ and $-Y$ indicate the rotation axes. We list the Hamiltonian in the rotating frame in each time interval. The Floquet Hamiltonian can be obtained by averaging the listed terms.}
    \label{fig:pre_floquet_xyz}
\end{figure}

First, we review Refs.~\cite{scholl2022microwave,geier2021floquet}, which initially prepare the {\it XY} Hamiltonian~\eqref{eq1}, and implement the {\it XYZ} Hamiltonian by periodically applying pulses. The pulse operation for implementing the {\it XYZ} model Hamiltonian is shown in Fig.~\ref{fig:pre_floquet_xyz}. It uses only four $\pi/2$ global rotations around the $\hat{S}^x, -\hat{S}^y, \hat{S}^y,$ and $-\hat{S}^x$ axes. In $\hat{H}_{\rm G}(t)$ of Eq.~\eqref{eq:exprsssion_of_effective_Hamiltonian}, this corresponds to setting the angle $\phi$ to $0, -\pi/2, \pi/2$, and $\pi$. The time intervals between pulses are given by $\tau_1, \tau_2$, and $2\tau_3$. Then, by using the Floquet Magnus expansion, the effective Hamiltonian of the system can be written by
\begin{align}
    \notag \hat{H}^{(0)}_{\rm{F}}=&\sum_{\langle i,j \rangle}\frac{2J_{ij}}{T_{\rm c}} \Big[(\tau_1+\tau_2)\hat{S}^{x}_{\bm{R}_{i}}\hat{S}^{x}_{\bm{R}_{j}}\\
    &+(\tau_1+\tau_3)\hat{S}^{y}_{\bm{R}_{i}}\hat{S}^{y}_{\bm{R}_{j}}
    +(\tau_2+\tau_3)\hat{S}^{z}_{\bm{R}_{i}}\hat{S}^{z}_{\bm{R}_{j}}\Big],
    \label{eq:pre_xyz}
\end{align}
where $T_{\rm c}=2(\tau_1+\tau_2+\tau_3)$ and $\sum_{\langle i,j\rangle}$ denotes the summation over the nearest-neighbor pairs.

\begin{figure}[H]
    \centering
    \includegraphics[width=8cm]{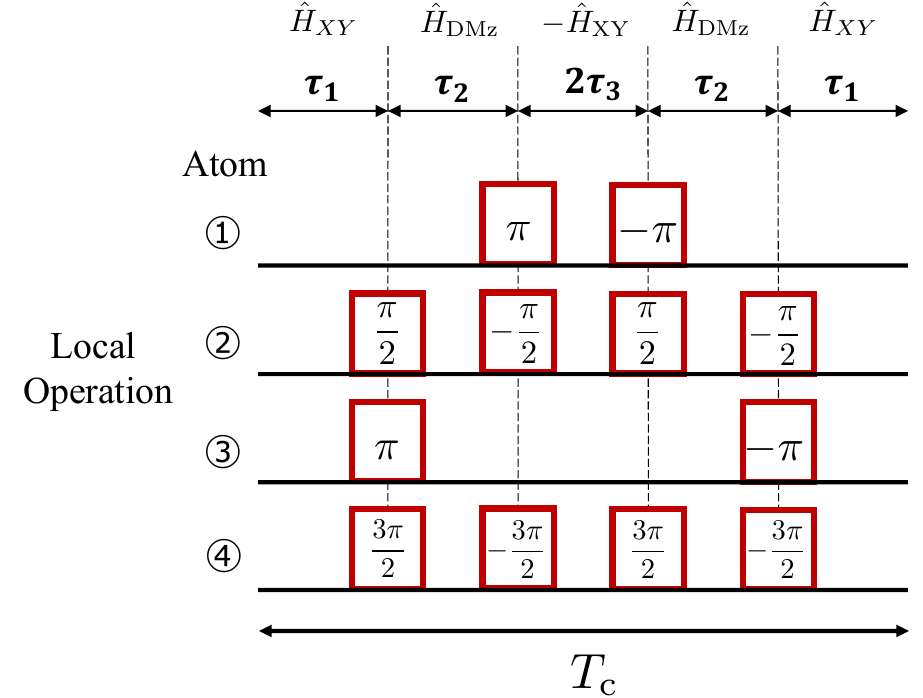}
    \caption{Pulse sequence for creating the Hamiltonian with the mono-axial DM interaction. Here we show a sequence with one cycle period, denoted by $T_{\rm c}$. Local operations refer to applying pulses to individual trapped atoms. The angles $\pm \pi/2, \pi, 3\pi/2$ indicate the rotation angles in the laboratory frame. All rotation axes are along $\hat{S}^z$-axis. We list the Hamiltonian of the rotating frame in each time interval. Here, we define $\hat{H}_{\rm{DM}z}\equiv\sum_{\langle i,j \rangle}D_{z}(\hat{S}^{x}_{\bm{R}_{i}}\hat{S}^{y}_{\bm{R}_{j}}-\hat{S}^{y}_{\bm{R}_{i}}\hat{S}^{x}_{\bm{R}_{j}})$.
    }
    \label{fig:pre_floquet_dmz}
\end{figure}

Second, we review Ref.~\cite{PhysRevA.108.053318}, which proposes a method for implementing the Hamiltonian with mono-axial DM interactions ($\bm{D}_{ij}\propto \bm{e}_z$) using local pulse operations. Figure~\ref{fig:pre_floquet_dmz} shows the proposed pulse sequence. Here, we consider the one-dimensional chain for simplicity. The pulse sequence consists of pulse operations so that the phase difference between neighboring atoms becomes $\pi/2$. By using the Floquet Magnus expansion, the effective Hamiltonian is obtained as
\begin{align}
    \notag \hat{H}^{(0)}_{\rm{F}}=&\sum_{\langle i,j \rangle} \frac{2J_{ij}}{T_{\rm c}}\Big[(\tau_1-\tau_3)(\hat{S}^{x}_{\bm{R}_{i}}\hat{S}^{x}_{\bm{R}_{j}}+\hat{S}^{y}_{\bm{R}_{i}}\hat{S}^{y}_{\bm{R}_{j}})\\
    &+\tau_2(\hat{S}^{x}_{\bm{R}_{i}}\hat{S}^{y}_{\bm{R}_{j}}-\hat{S}^{y}_{\bm{R}_{i}}\hat{S}^{x}_{\bm{R}_{j}})\Big],
    \label{eq:pre_mono_dm}
\end{align}
where $\bm{R}_j$ represents the one-dimensional lattice position. 

\section{Results}\label{sec:results}
In two dimensions, various types of DM interactions can exist. In this work, we propose a method to create Bloch-type and N\'{e}el-type DM interactions~\cite{Derras-Chouk2018-hc,Mahfouzi2021-uq} in addition to Heisenberg interactions. Before presenting our results, we explain our strategy for implementing these interactions using Floquet engineering.

The DM vectors of Bloch- and N\'{e}el-type DM interactions are given by $\bm{D}_{ij}\propto \bm{e}_x$ and $\propto\bm{e}_y$, respectively [see Eqs.~(\ref{eq:target_Hamiltonian_Bloch_type}) and (\ref{eq:target_Hamiltonian_Neel})]. The key point for creating these terms is that the DM interactions can be obtained by applying a local spin rotation to the interaction terms $\hat{h}_{ij}^{zx}\equiv \hat{S}_{\bm{R}_i}^z\hat{S}_{\bm{R}_j}^z+\hat{S}_{\bm{R}_i}^x\hat{S}_{\bm{R}_j}^x$ and $\hat{h}_{ij}^{yz}\equiv \hat{S}_{\bm{R}_i}^y\hat{S}_{\bm{R}_j}^y+\hat{S}_{\bm{R}_i}^z\hat{S}_{\bm{R}_j}^z$. For example, $\hat{h}_{ij}^{zx}$ is transformed to $\hat{S}_{\bm{R}_i}^z\hat{S}_{\bm{R}_j}^z-\hat{S}_{\bm{R}_i}^x\hat{S}_{\bm{R}_j}^x$ after local spin rotation around the $\hat{S}^y$ axis. For details on the relation between site-dependent spin rotation and the
DM interaction, see Appendix~\ref{sec:appendix_rotation}. Since the terms $\hat{h}_{ij}^{zx}$ and $\hat{h}_{ij}^{yz}$ appear after global $\pi/2$ spin rotation of the {\it XY} Hamiltonian around the $\hat{S}^x$ or $\hat{S}^y$ axis (see Fig.~\ref{fig:pre_floquet_xyz}), we can obtain Bloch- and N\'{e}el-type DM interactions combining the global and local pulse operations appropriately. The detailed methods for obtaining these two types of DM interactions are provided in Secs.~\ref{Bloch-type DM int.} and \ref{Neel-type DM int.}. 

The resultant pulse sequence consists of 14 time intervals, from $\tau_1$ to $\tau_{14}$ (see Figs.~\ref{fig: Bloch type pulse} and \ref{fig: Neel type pulse}). We note that the validity of the Floquet Magnus expansion is determined by the total period of the pulse sequence $T$, which is given by the sum of all $\tau_j$ rather than by each interval $\tau_j$. For a typical interaction strength of the Rydberg atoms $J_0\simeq h\times 250~{\rm kHz}$, the validity condition for the Floquet Magnus expansion $J_0T/\hbar\ll 2\pi$ can be satisfied for $T\lesssim 1~\mu${\rm s}, which is feasible with current experimental techniques~\cite{geier2021floquet,scholl2022microwave}. 

In the main text, we focus on the nearest-neighbor interactions between Rydberg atoms for simplicity (see Appendix~\ref{sec:appendix_derivation_Bloch_and_Neel} for a detailed derivation). A discussion of long-range interaction, including next-nearest and next-next-nearest neighbors, is provided in Appendix~\ref{sec:long-range}. 

\subsection{Bloch-type DM interaction}\label{Bloch-type DM int.}
\begin{figure*}[t]
    \centering
    \includegraphics[width=\textwidth]{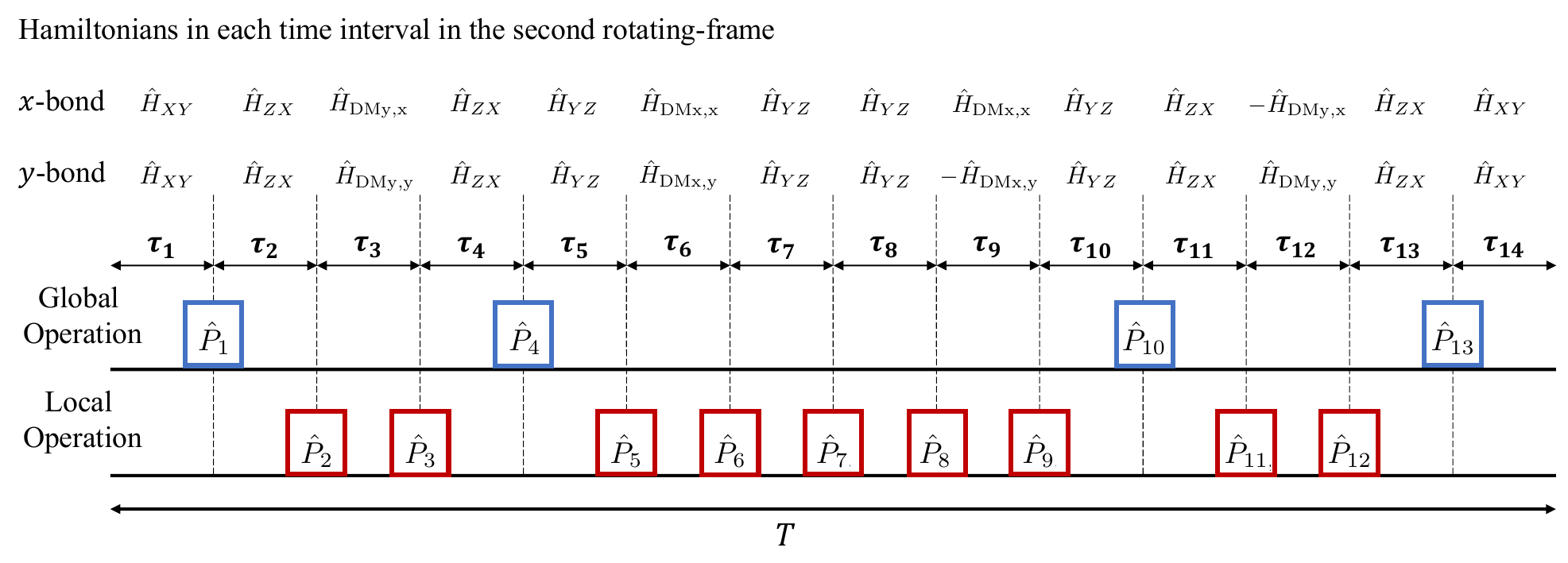}
    \caption{Proposed pulse sequence for creating the Bloch-type DM interaction. Here, we show the sequence of a single period $T$. Global and local operations refer to applying the pulses to whole atoms and individual atoms, respectively. The operator $\hat{P}_{i}$ denotes the pulse operator in the laboratory frame at time $T_j$. We list the Hamiltonian of the rotating frame in each time interval for each bond. Here, we define $\hat{H}_{\rm{DM}\alpha,{\beta}}\equiv\sum_{i}D_{\alpha}(\hat{\bm{S}}_{\bm{R}_{i}}\times\hat{\bm{S}}_{\bm{R}_{i}+a\bm{e}_{\beta}}){}_{\alpha}$.}
    \label{fig: Bloch type pulse}
\end{figure*}
First, we consider the Bloch-type DM interaction. The target Hamiltonian is given by
\begin{align}
    \notag \hat{H}^{\mathrm{B}}_{\mathrm{target}}=\sum_{\langle i,j \rangle}J^{\rm ex}_{ij}&(\hat{\bm{S}}_{\bm{R}_{i}}\cdot\hat{\bm{S}}_{\bm{R}_{j}})\\
    \notag +\sum_{i}\Big[&D_x(\hat{\bm{S}}_{\bm{R}_{i}}\times\hat{\bm{S}}_{\bm{R}_{i}+a\bm{e}_x}){}_x\\
    +&D_y(\hat{\bm{S}}_{\bm{R}_{i}}\times\hat{\bm{S}}_{\bm{R}_{i}+a\bm{e}_y}){}_y\Big],\label{eq:target_Hamiltonian_Bloch_type}
\end{align}
where $J_{ij}^{\rm ex}$ is the strength of the exchange interaction, and $D_x$ and $D_y$ represent the strength of the DM interaction along the $x$ bond and $y$ bond, respectively [see Fig.~\ref{fig:schematic diagram}(a)]. The term $(\hat{\bm{S}}_{\bm{R}_i}\times\hat{\bm{S}}_{\bm{R}_j})_{\alpha}$ denotes the $\alpha$ component of the outer product. This Hamiltonian can be realized by combining global and local operations within a single period. Our proposed pulse sequence is shown in Fig.~\ref{fig: Bloch type pulse}. After applying Floquet engineering, we obtain the following effective Hamiltonian (see Appendix~\ref{sec:appendix_derivation_Bloch_and_Neel} for details):
\begin{align}
    \notag \hat{H}^{(0)}_{\mathrm{F}}&=\alpha_{XY}\hat{H}_{XY}+\alpha_{YZ}\hat{H}_{YZ}+\alpha_{XZ}\hat{H}_{ZX}\\
    \notag &+\sum_{i}\frac{J}{T}\Big[(\tau_6+\tau_9)(\hat{\bm{S}}_{\bm{R}_{i}}\times\hat{\bm{S}}_{\bm{R}_{i}+a\bm{e}_x}){}_x\\
    \notag&+(\tau_
    3-\tau_{12})(\hat{\bm{S}}_{\bm{R}_{i}}\times\hat{\bm{S}}_{\bm{R}_{i}+a\bm{e}_x}){}_y\\
    \notag&+(\tau_6-\tau_9)(\hat{\bm{S}}_{\bm{R}_{i}}\times\hat{\bm{S}}_{\bm{R}_{i}+a\bm{e}_y}){}_x\\
    &+(\tau_3+\tau_{12})(\hat{\bm{S}}_{\bm{R}_{i}}\times\hat{\bm{S}}_{\bm{R}_{i}+a\bm{e}_y}){}_y \Big],
\end{align}
where $J\equiv 2C_3/a^3$ denotes the strength of the nearest-neighbor bare interaction, $\alpha_{XY}\equiv(\tau_1+\tau_{14})/T$, $\alpha_{YZ}\equiv(\tau_5+\tau_7+\tau_8+\tau_{10})/T$, $\alpha_{ZX}\equiv(\tau_2+\tau_4+\tau_{11}+\tau_{13})/T$, and we defined the following Hamiltonians:
\begin{align}
    \hat{H}_{XY}=J\sum_{\langle i,j \rangle}(\hat{S}^{x}_{\bm{R}_{i}}\hat{S}^{x}_{\bm{R}_{j}}+\hat{S}^{y}_{\bm{R}_{i}}\hat{S}^{y}_{\bm{R}_{j}}),\\
    \hat{H}_{YZ}=J\sum_{\langle i,j \rangle}(\hat{S}^{y}_{\bm{R}_{i}}\hat{S}^{y}_{\bm{R}_{j}}+\hat{S}^{z}_{\bm{R}_{i}}\hat{S}^{z}_{\bm{R}_{j}}),\\
    \hat{H}_{ZX}=J\sum_{\langle i,j \rangle}(\hat{S}^{z}_{\bm{R}_{i}}\hat{S}^{z}_{\bm{R}_{j}}+\hat{S}^{x}_{\bm{R}_{i}}\hat{S}^{x}_{\bm{R}_{j}}).
\end{align}
To implement the target Hamiltonian (\ref{eq:target_Hamiltonian_Bloch_type}), we set the propagation times as $\tau_1+\tau_{14}=\tau_2+\tau_4+\tau_{11}+\tau_{13}=\tau_5+\tau_7+\tau_8+\tau_{10}\equiv\tau$, $\tau_6=\tau_9\equiv\tau_x$, and $\tau_3=\tau_{12}\equiv\tau_y$. 
The effective Hamiltonian then becomes
\begin{align}
    \notag \hat{H}^{(0)}_{\rm{F,Bloch}}&=J_{\rm{F}}\sum_{\langle i,j\rangle}\hat{\bm{S}}_{\bm{R}_{i}}\cdot\hat{\bm{S}}_{\bm{R}_{j}}\\
    \notag&+\sum_i\Big[D_{x,\rm{F}}(\hat{\bm{S}}_{\bm{R}_{i}}\times\hat{\bm{S}}_{\bm{R}_{i}+a\bm{e}_x}){}_x\\
    &+D_{y,\rm{F}}(\hat{\bm{S}}_{\bm{R}_{i}}\times\hat{\bm{S}}_{\bm{R}_{i}+a\bm{e}_y}){}_y\Big],
\end{align}
where we have defined $J_{\rm{F}}=J\tau/T$, $D_{x,\rm{F}}=J\tau_x/T$, and $D_{y,\rm{F}}=J\tau_y/T$. We note that the period $T$ can be written as $T=\sum_{k=1}^{14}\tau_k$. Although the ratio between the exchange interaction and DM interaction $J/D$ can be tuned  from $0$ to $\infty$ by changing the time interval, there are practical limitations of tuning the ratio $J/D$ due to the experimental constraints, such as the finite pulse width. 

Here, we show the pulse sequence for creating the Bloch-type DM interaction. Let $\hat{P}_{i}$ be a unitary operator representing the time evolution at time $T_i\equiv \sum_{k=1}^i\tau_k$. This operator describes the time evolution in the laboratory frame. The expressions are given by (see Appendix~\ref{sec:appendix_derivation_Bloch_and_Neel} for details)
\begin{align}
    &\hat{P}_{1}=\exp(-i\hat{S}^{x}_{\rm{tot}}\pi/2),\\
    &\hat{P}_{2}=\exp\Big\{-i\sum_{j}\hat{S}^{z}_{\bm{R}_{j}}[2\pi-(n_j+m_j)\pi/2]\Big\},\\
    &\hat{P}_{3}=\hat{P}^{\dagger}_{2},\\
    &\hat{P}_{4}=\exp(+i\hat{S}^{y}_{\rm{tot}}\pi/2),
\end{align}
\begin{align}
    &\hat{P}_{5}=\hat{P}_{2},\\
    &\hat{P}_{6}=\hat{P}_{3},\\
    &\hat{P}_{7}=\hat{1},\\
    &\hat{P}_{8}=\exp\Big\{-i\sum_{j}\hat{S}^{z}_{\bm{R}_{j}}[2\pi+(m_j-n_j)\pi/2]\Big\},\\
    &\hat{P}_{9}=\hat{P}^{\dagger}_{8},\\
    &\hat{P}_{10}=\hat{P}^{\dagger}_{4},\\
    &\hat{P}_{11}=\exp\Big\{-i\sum_{j}\hat{S}^{z}_{\bm{R}_{j}}[2\pi+(n_j-m_j)\pi/2]\Big\},\\
    &\hat{P}_{12}=\hat{P}^{\dagger}_{11},\\
    &\hat{P}_{13}=\hat{P}^{\dagger}_{1},
\end{align}
where $\hat{S}^{\alpha}_{\rm tot}=\sum_j\hat{S}^{\alpha}_{\bm{R}_j}$, $\hat{1}$ denotes the identity operator, and $n_j$ ($m_j$) represent the $x (y)$-coordinates of atoms on a two-dimensional lattice.

\subsection{N\'{e}el-type DM interaction}\label{Neel-type DM int.}
\begin{figure*}[t]
    \centering
    \includegraphics[width=\textwidth]{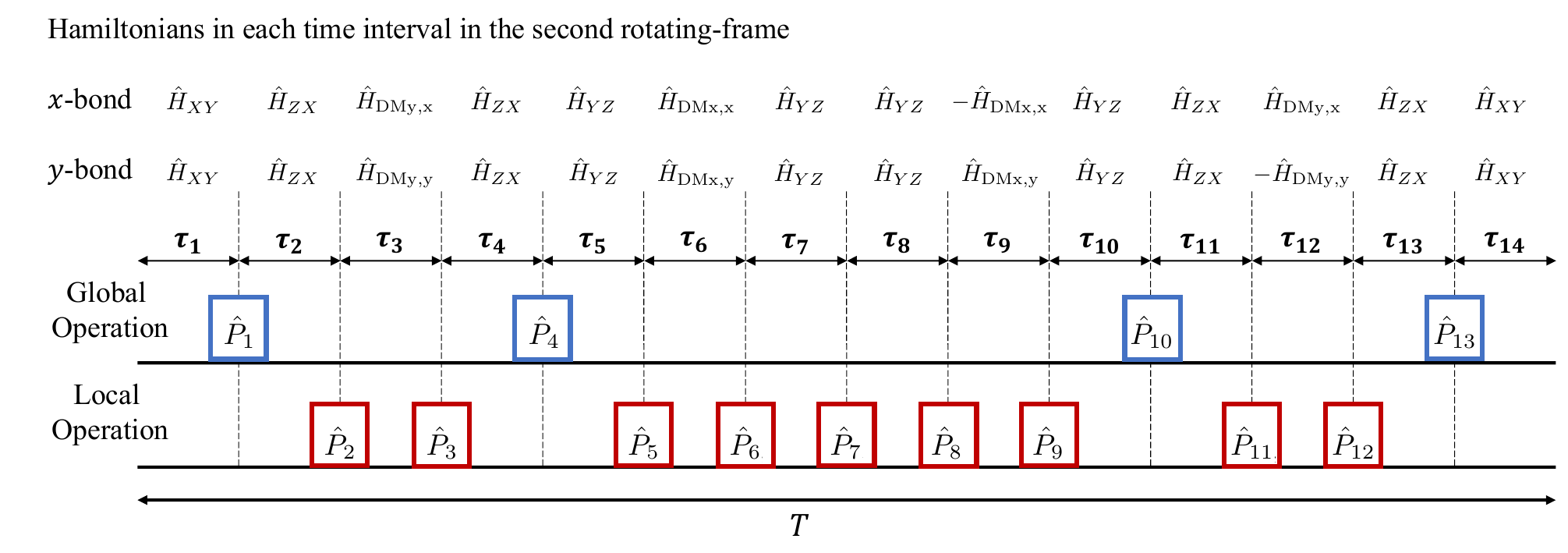}
    \caption{Proposed pulse sequence for creating the N\'{e}el type DM interaction. Here, we show the sequence of a single period $T$. Global and local operations refer to applying the pulses to whole atoms and individual atoms, respectively. The operator $\hat{P}_{i}$ denotes the pulse operator in the laboratory frame at time $T_j$. We list the Hamiltonian of the rotating frame in each time interval for each bond.}
    \label{fig: Neel type pulse}
\end{figure*}
Next, we discuss how to create the N\'{e}el-type DM interaction. The target Hamiltonian is given by
\begin{align}
    \notag \hat{H}^{\mathrm{N}}_{\mathrm{target}}=\sum_{\langle i,j \rangle}J_{ij}&(\hat{\bm{S}}_{\bm{R}_{i}}\cdot\hat{\bm{S}}_{\bm{R}_{j}})\\
    \notag +\sum_{i}\Big[&D_y(\hat{\bm{S}}_{\bm{R}_{i}}\times\hat{\bm{S}}_{\bm{R}_{i}+a\bm{e}_x}){}_y\\
    +&D_x(\hat{\bm{S}}_{\bm{R}_{i}}\times\hat{\bm{S}}_{\bm{R}_{i}+a\bm{e}_y}){}_x\Big].
    \label{eq:target_Hamiltonian_Neel}
\end{align}
This Hamiltonian can be realized by slightly modifying the pulse sequence used for the Bloch-type DM interaction. We replace $\hat{P}_{8}$ and $\hat{P}_{11}$ with 
\begin{align}
    \hat{P}_{8}&=\exp\Big\{-i\sum_{j}\hat{S}^{z}_{\bm{R}_{j}}[2\pi-(m_j-n_j)\pi/2]\Big\},\\
    \hat{P}_{11}&=\exp\Big\{-i\sum_{j}\hat{S}^{z}_{\bm{R}_{j}}[2\pi-(n_j-m_j)\pi/2]\Big\}.
\end{align}
Using the pulse sequence shown in Fig.~\ref{fig: Neel type pulse}, we obtain the effective Hamiltonian
\begin{align}
    \notag \hat{H}^{(0)}_{\mathrm{F}}&=\alpha_{XY}\hat{H}_{XY}+\alpha_{YZ}\hat{H}_{YZ}+\alpha_{XZ}\hat{H}_{ZX}\\
    \notag &+\sum_{i}\frac{J}{T}\left[(\tau_6-\tau_9)(\hat{\bm{S}}_{\bm{R}_{i}}\times\hat{\bm{S}}_{\bm{R}_{i}+a\bm{e}_x}){}_x\right.\\
    \notag&+(\tau_3+\tau_{12})(\hat{\bm{S}}_{\bm{R}_{i}}\times\hat{\bm{S}}_{\bm{R}_{i}+a\bm{e}_x}){}_y\\
    \notag&+(\tau_6+\tau_9)(\hat{\bm{S}}_{\bm{R}_{i}}\times\hat{\bm{S}}_{\bm{R}_{i}+a\bm{e}_y}){}_x\\&\left.+(\tau_3-\tau_{12})(\hat{\bm{S}}_{\bm{R}_{i}}\times\hat{\bm{S}}_{\bm{R}_{i}+a\bm{e}_y}){}_y \right].
\end{align}
Setting the pulse interval appropriately, we obtain the target Hamiltonian Eq.~\eqref{eq:target_Hamiltonian_Neel}.
\section{Numerical Analysis}\label{sec:numerical analysis}

\begin{figure*}[t]
    \centering
    \includegraphics[width=\textwidth]{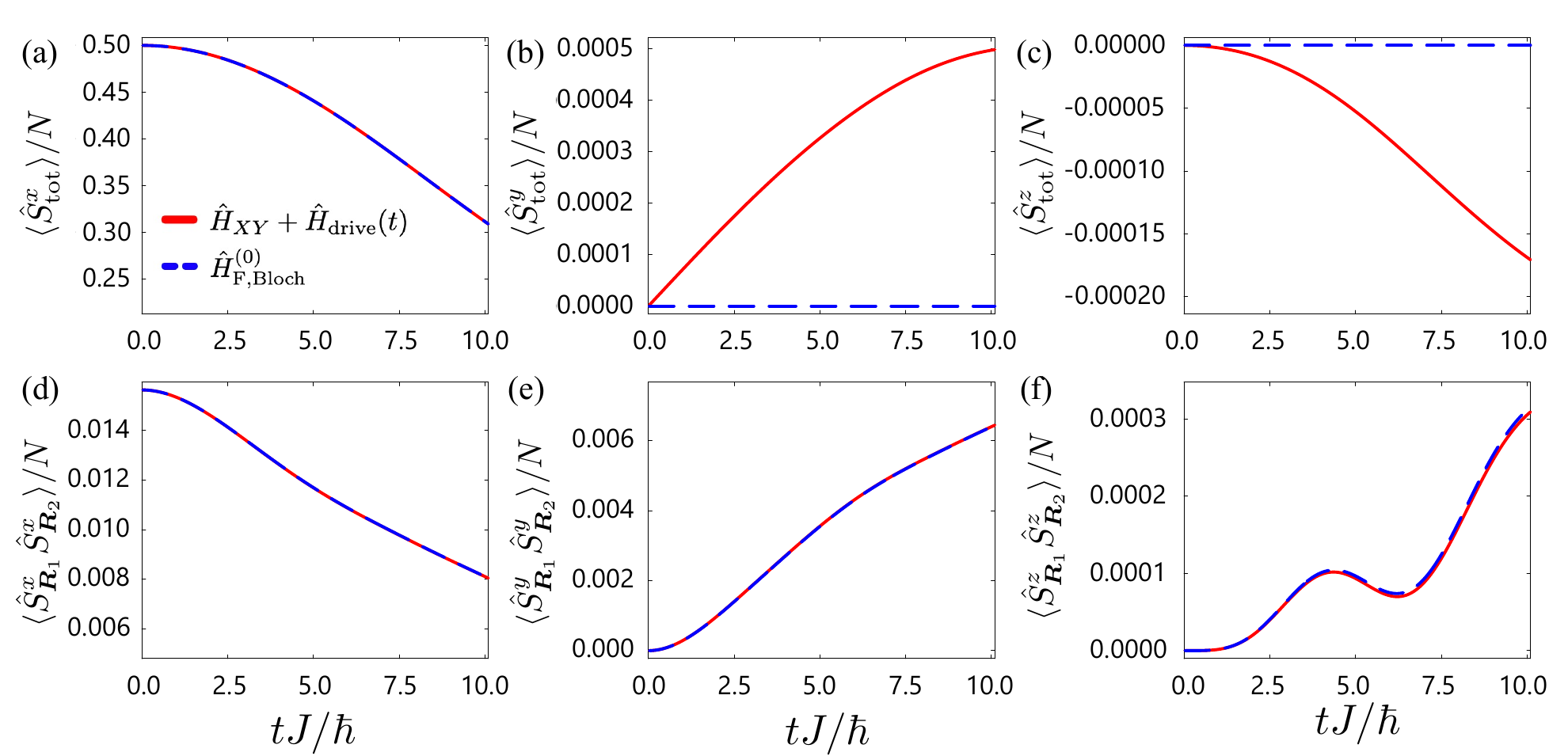}\\
    \caption{Time evolution of the spin expectation value $\langle\hat{S}^{\alpha}_{\rm tot}\rangle/N$ for $(a)\;\alpha=x$, $(b)\;\alpha=y$, and $(c)\;\alpha=z$. The time evolution of the two-body correlation function $\langle\hat{S}^{\alpha}_{\bm{R}_1}\hat{S}^{\alpha}_{\bm{R}_2}\rangle/N$ for $(d)\;\alpha=x$, $(e)\;\alpha=y$, and $(f)\;\alpha=z$. The solid red (dashed blue) line represents the results of the time-evolution under the original Hamiltonian $\hat{H}_{XY}+\hat{H}_{\rm drive}(t)$ (the Bloch-type effective Hamiltonian $\hat{H}_{\rm F,Bloch}^{(0)}$)}
    \label{fig: num result}
\end{figure*}

In this section, we show numerical results for the time evolution under the original Hamiltonian $\hat{H}(t)=\hat{H}_{XY}+\hat{H}_{\rm drive}(t)$ and effective Hamiltonian $\hat{H}_{\rm F, Bloch}^{(0)}$ to verify the validity of our scheme. We calculate the spin expectation value $\langle\hat{S}^{\alpha}_{\rm rot}\rangle/N$ and the two-body correlation function $\langle\hat{S}^{\alpha}_{\bm{R}_1}\hat{S}^{\alpha}_{\bm{R}_2}\rangle/N$, where $N$ is the number of lattice sites. The initial condition is set to $\ket{\psi(t=0)}\equiv\prod_{j=1}^{N}\ket*{+_j},\ket*{+_j}\equiv(\ket{\uparrow_j}+\ket{\downarrow_j})/\sqrt{2}$. We assume a $4\times 4$ square lattice. The other parameters are set to $T=0.028\hbar/J$, $\tau_1=\tau_{14}=0.002 (\hbar/J), \tau_2=\tau_4=\tau_5=\tau_7=\tau_8=\tau_{10}=\tau_{11}=\tau_{13}=0.001 (\hbar/J)$, and $\tau_3=\tau_6=\tau_9=\tau_{12}=0.004 (\hbar/J)$. Under this setting, we obtain $J_{\rm F}=D_{x,{\rm F}}=D_{y,{\rm F}}\simeq 0.14J$. The numerical calculations are performed using the exact diagonalization method and the \texttt{exponentiate} function of the KrylovKit.jl software package~\cite{krylovkit}.

\subsection{Ideal case}

We show the numerical results of the spin expectation value in Figs.~\ref{fig: num result}(a)-\ref{fig: num result}(c). The time evolution of $\hat{S}^x_{\rm tot}$ under the effective Hamiltonian is in good agreement with that under the original Hamiltonian as shown in Fig.~\ref{fig: num result}(a). However, we find the clear deviation for $\hat{S}^y_{\rm tot}$ and $\hat{S}^z_{\rm tot}$ as shown in Figs.~\ref{fig: num result}(b) and \ref{fig: num result}(c). 
Under the time evolution associated with the effective Hamiltonian, the expectation
values of $\hat{S}^y_{\rm tot}$ and $\hat{S}^z_{\rm tot}$ are always zero at any time. 
These results are due to a symmetry of the effective Hamiltonian (see Appendix~\ref{sec:spin expectation values} for details). The exact Hamiltonian has no such symmetry possessed by the effective Hamiltonian. Therefore, the expectation values of $\hat{S}_{\rm tot}^y$ and $\hat{S}_{\rm tot}^z$ deviate from zero in the exact time evolution. If we consider the higher-order term of the Magnus expansion, the Floquet Hamiltonian might describe this deviation. In Figs.~\ref{fig: num result}(d)-\ref{fig: num result}(f), we show the time evolution of the correlation functions $\langle\hat{S}_{\bm{R}_1}^{\alpha}\hat{S}_{\bm{R}_2}^{\alpha}\rangle/N\; (\alpha=x,y,z)$, where $\bm{R}_1=(0,0)$ and $\bm{R}_2=(a,0)$. We can see good agreement with the exact and effective Hamiltonian. From these results, we conclude that our proposed pulse scheme can successfully implement the effective Hamiltonian.

\begin{figure*}[t]
    \centering
    \includegraphics[width=\textwidth]{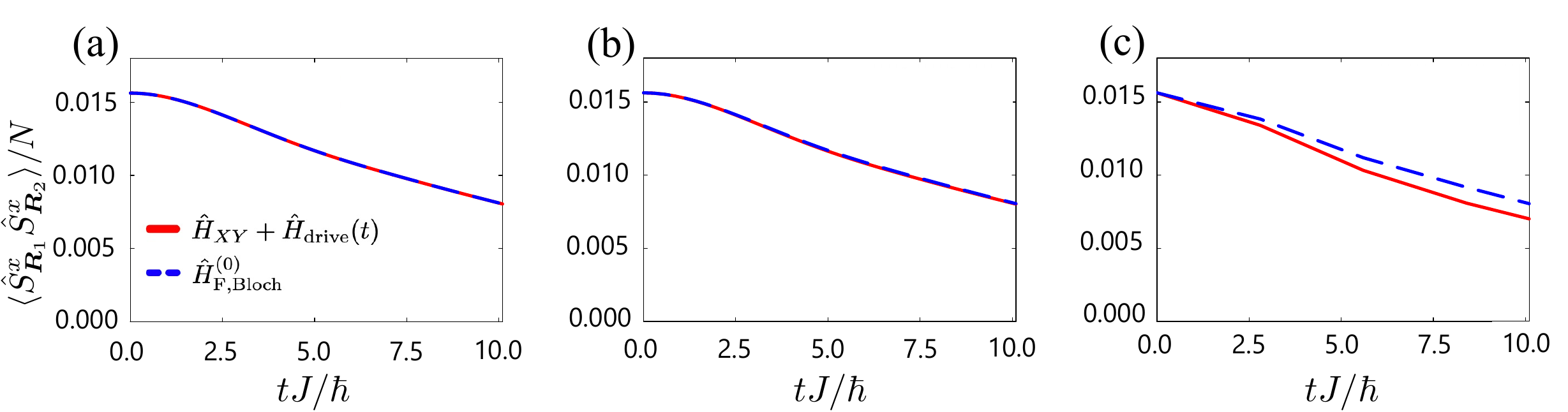}
    \caption{Time evolution of the two-body correlation function $\langle\hat{S}^{x}_{\bm{R}_1}\hat{S}^{x}_{\bm{R}_2}\rangle/N$ for different periods $T$: (a) $T=0.028\hbar/J$, (b) $T=0.28\hbar/J$, and (c) $T=2.8\hbar/J$. The red solid line shows the time evolution under the original Hamiltonian $\hat{H}_{XY}+\hat{H}_{\rm drive}(t)$, while the blue dashed line represents the results under the Bloch-type effective Hamiltonian $\hat{H}_{\rm F,Bloch}^{(0)}$.}
    \label{fig:T_compare}
\end{figure*}

Finally, we discuss the validity of the lowest-order Magnus expansion in our systems. As discussed in Sec.~\ref{sec:model} and the previous paragraph of this section, the Magnus expansion is a high-frequency expansion. The accuracy of the method depends on the period $T$ of the time-periodic external field. To see explicitly how the period $T$ affects the validity of the expansion, we calculate the time evolution of the correlation function $\langle\hat{S}^{x}_{\bm{R}_1}\hat{S}^{x}_{\bm{R}_2}\rangle/N$ for the same interaction parameters as Fig.~\ref{fig: num result}, but for different periods $T$. The results are shown in Figs.~\ref{fig:T_compare}(a)-\ref{fig:T_compare}(c). In the short-period regime ($JT/\hbar\lesssim 2\pi$), the lowest-order Magnus expansion agrees well with the exact results as shown in Figs.~\ref{fig:T_compare}(a) and \ref{fig:T_compare}(b). On the other hand, in the long-period ($JT/\hbar\sim 2\pi$) case, we can see a clear deviation between the lowest-order Magnus expansion and exact results for $T=2.8\hbar/J$, as shown in Fig.~\ref{fig:T_compare}(c). This discrepancy indicates that, for longer periods, it is necessary to incorporate higher-order contributions in the Magnus expansion.

\subsection{Realistic case}
\begin{figure}[h]
    \centering
    \includegraphics[width=8cm]{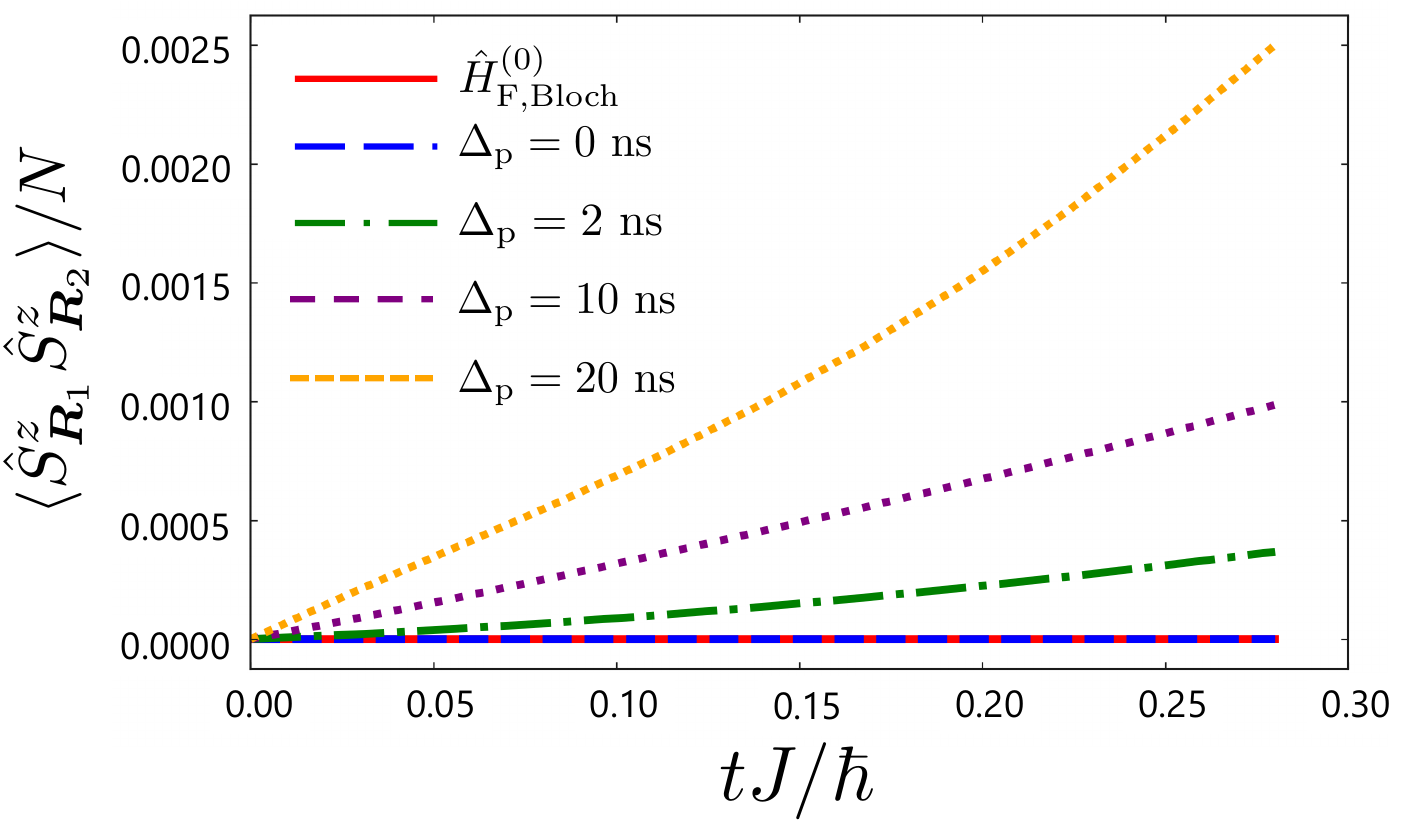}
    \caption{Time evolution of the two-body correlation function $\langle\hat{S}^{z}_{\bm{R}_1}\hat{S}^{z}_{\bm{R}_2}\rangle/N$. The blue dashed, green dash-dotted, purple dotted (wide spacing), and yellow dotted (narrow spacing) lines show the time evolution under the original Hamiltonian $\hat{H}_{XY}+\hat{H}_{\rm drive}(t)$ with finite pulse width of $\Delta_{\rm p}=0, 2, 10, \text{and } 20~{\rm ns}$, respectively, while the red solid line represents the result under the Bloch-type effective Hamiltonian $\hat{H}_{\rm F,Bloch}^{(0)}$.}
    \label{fig:corr_z_finite_pulse}
\end{figure}
We now turn to discuss the experimental imperfections. When implementing the Hamiltonian via Floquet engineering on a Rydberg‐atom quantum simulator, the main sources of imperfection are the finite lifetime of the Rydberg atoms and the finite width of the driving pulses.

First, we discuss decoherence due to spontaneous emission from the Rydberg state. Previous studies have experimentally confirmed lifetimes on the order of $\mathcal{O}(10\text{–}100\rm{\mu s})$~\cite{PhysRevLett.124.023201,PhysRevA.110.043114,browaeys2020many}. In contrast, in our numerical calculations we assume a period of $1.8~\mu$s per cycle, which is shorter than the current decoherence times of the Rydberg atoms. Therefore, in this work, we treat the finite pulse width of the control pulses as the primary experimental imperfection.

Furthermore, using numerical simulations, we analyze two cases: an idealized case with $\Delta_{\rm p} = 0$ and a more practical case involving finite-width Gaussian pulses of the form $\frac{1}{\sqrt{2\pi}\,\Delta_{\rm p}}\exp\!\Bigl[-\frac{(t - \tau_i)^2}{2\,\Delta_{\rm p}^2}\Bigr]$, where $\Delta_{\rm p}$ represents the pulse width, as shown in Fig.~\ref{fig:corr_z_finite_pulse}. 
\begin{figure}[h]
    \centering
    \includegraphics[width=8cm]{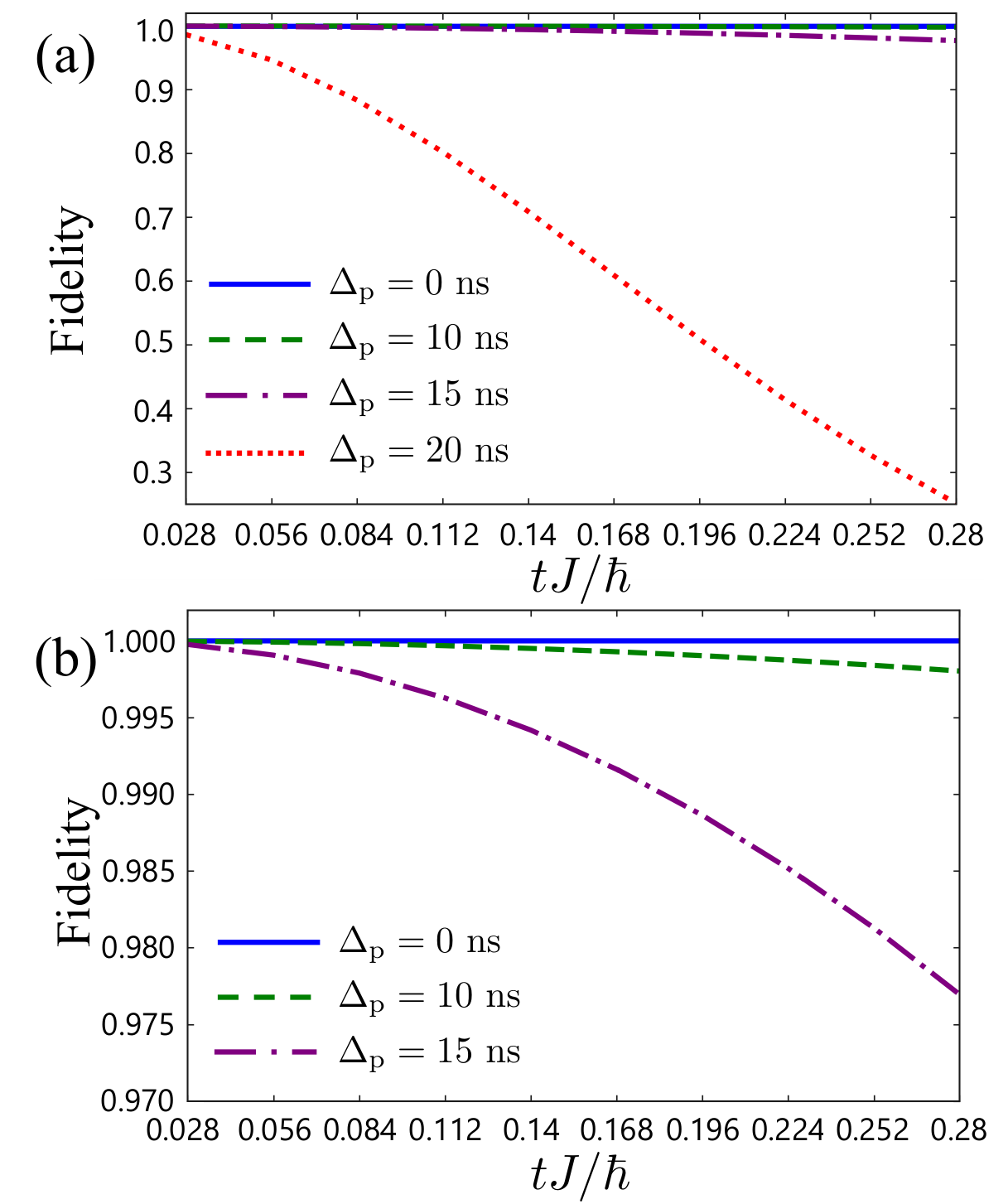}
    \caption{(a) Fidelity per Floquet cycle $T=0.028\hbar/J$ between the state evolved under the effective Hamiltonian $\hat{H}_{\rm F,Bloch}^{(0)}$ and the state evolved under the original Hamiltonian $\hat{H}_{XY}+\hat{H}_{\rm drive}(t)$ using finite-width Gaussian pulses with durations $\Delta_{\rm p} = 0, 10, 15$, and $20$ ns. The blue solid, green dashed, purple dash-dotted, and red dotted lines correspond to $\Delta_{\rm p} = 0, 10, 15$ and $\Delta_{\rm p} = 20$ ns, respectively. (b) A magnified view of the high-fidelity regime in (a).
    }
    \label{fig:fidelity}
\end{figure}

In addition to examining the time evolution of physical observables under finite pulse widths, we also investigate the impact on the time-evolution operator itself.
Let $\hat{U}_{\mathrm F}$ denote the time-evolution operator in the ideal case, governed by the effective Hamiltonian $\hat{H}^{(0)}_{\mathrm {F,Bloch}}$, and let $\hat{U}_{\Delta_{\rm p}}$ denote the time-evolution operator under finite-width Gaussian pulses with a pulse width $\Delta_{\rm p}$. To quantify the influence of the pulse width on the time-evolution operator, we evaluate the temporal variation of the fidelity ${|\bra{\psi(t=0)}\hat{U}^{\dagger}_{\mathrm F}\hat{U}_{\Delta_{\rm p}}\ket{\psi(t=0)}|}^2$, as shown in Fig~\ref{fig:fidelity}.

\begin{figure}[h]
    \centering
    \includegraphics[width=8cm]{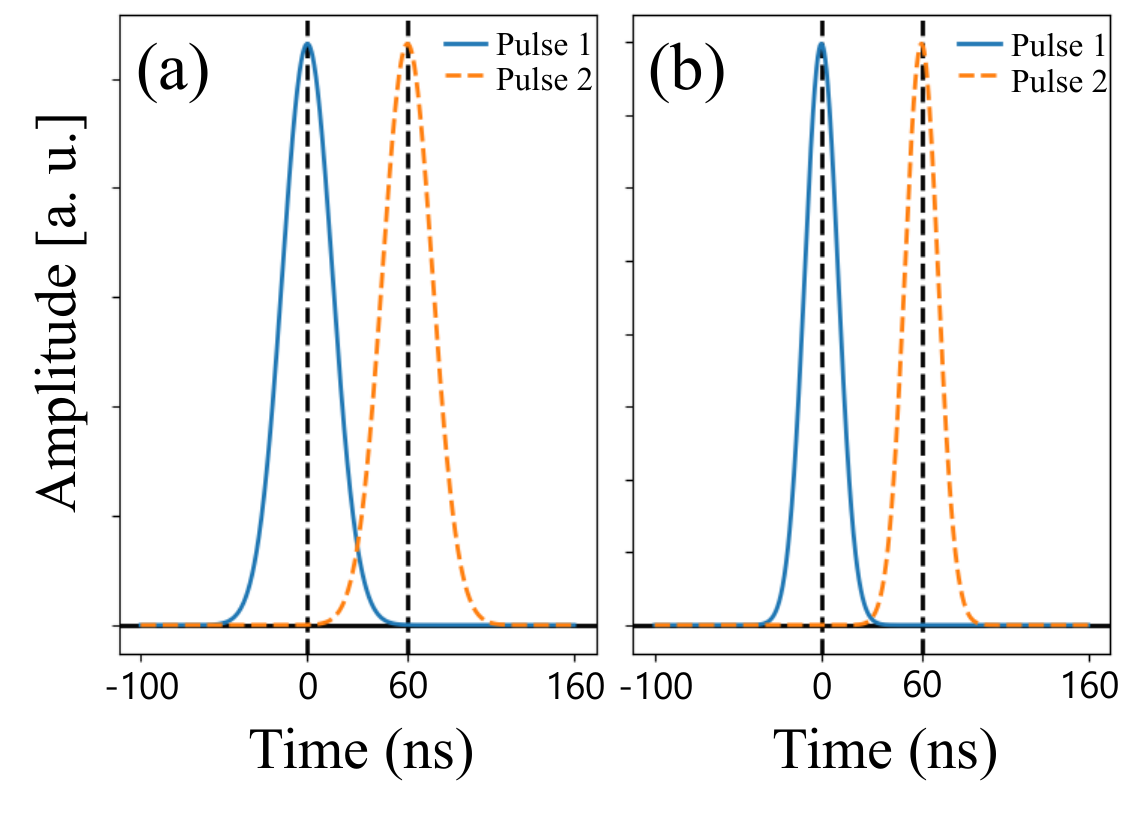}
    \caption{Gaussian pulses with widths of (a) $20~$ns and (b) $10~$ns, shown within the shortest free‐evolution interval of 60 ns between $\tau_i$ and $\tau_{i+1}$.}
    \label{fig:gaussian}
\end{figure}
From Fig.~\ref{fig:fidelity}, with a pulse width of $10~$ns, the fidelity at the tenth cycle is approximately $0.998$, corresponding to an error of only about $0.2$\%. This indicates that the pulse sequence proposed in this study reproduces the dynamics of the target Hamiltonian with high accuracy. In contrast, for a pulse width of $20~$ns, the fidelity at the tenth cycle drops to approximately $0.250$, resulting in a very large error. This degradation can be attributed to the fact that the shortest free-evolution interval between $\tau_i$ and $\tau_{i+1}$ in the proposed sequences (Figs.~\ref{fig: Bloch type pulse} and \ref{fig: Neel type pulse}) is on the order of $60~$ns; the increased pulse width therefore shortens the actual time available for evolution under the intended Hamiltonian. Moreover, as the pulse width increases, the overlap between adjacent pulses becomes non-negligible, leading to unintended interference and reduced fidelity in the simulated dynamics. We numerically plotted Gaussian pulses of width $\Delta_{\rm p}=20~$ns shown in Fig.~\ref{fig:gaussian}~(a) and $\Delta_{\rm p}=10~$ns in Fig.~\ref{fig:gaussian}~(b), within the shortest free‐evolution interval of $60~$ns between $\tau_i$ and $\tau_{i+1}$. Hence, for experimental implementation, it is necessary to carefully adjust both the pulse width used for atomic control and the durations of the free-evolution intervals.

\section{Summary}\label{sec:summary}
In conclusion, we have proposed a method to create the effective Hamiltonian incorporating the Heisenberg interactions and two-dimensional DM interactions through Floquet engineering. This approach utilizes both local and global operations tailored for Rydberg atoms arranged in a two-dimensional square lattice. The strength of each interaction in the proposed Hamiltonian can be controlled by tuning the pulse intervals. Numerical simulations have validated the effectiveness of our proposed sequence. These simulations cover both idealized cases and those based on experimentally attainable parameters~\cite{scholl2022microwave,de2017optical}, demonstrating that our proposal can be readily implemented with existing experimental techniques.

All pulse operations in our method can be feasibly implemented using state-of-the-art experimental techniques. 
Indeed, global operations~\cite{geier2021floquet,scholl2022microwave} and local operations~\cite{PhysRevA.108.053318,PhysRevLett.132.263601,de2017optical} have each been demonstrated experimentally. \textcolor{black}{Here, we comment on the number of the lasers for local operations. At first glance, the number of the lasers appears to scale linearly with the number of atoms. In optical tweezer research, this problem is not a bottleneck of the experiments~\cite{young2020half,norcia2018microscopic,PhysRevLett.132.263601,chen2023continuous,chen2023spectroscopy}. By applying a single laser onto a spatial light modulator and splitting the beam, one can achieve site‐selective addressing even in large‐scale systems.}

In this study, we implemented a Hamiltonian with Heisenberg and DM interactions on a two-dimensional square lattice. This work represents a step toward the future realization of quantum simulations of arbitrary quantum spin systems. We expect that the implemented Hamiltonian can be utilized to explore novel quantum many-body phenomena and contribute to the investigation of the microscopic behavior of quantum skyrmion by accessing regimes where quantum fluctuations are significant. 
Previous studies have shown that the skyrmion size depends on the ratio $J/D$, where $J$ is the strength of the exchange interaction and $D$ is the strength of the DM interaction. In particular, increasing $J$ leads to an enlargement of the skyrmion size, whereas increasing $D$ causes the skyrmion to shrink~\cite{PhysRevB.92.214439,Stepanov2019heisenberg,wang2018theory}. The Floquet engineering scheme we employ can access the large-$D$ limit.This tunability enables the realization of compact quantum skyrmions even in modest system sizes, such as $2 \times 2$ to $6 \times 6$ arrays, which are within the reach of current Rydberg atom experiments~\cite{Sotnikov2018,Stepanov2019heisenberg,Mland2022,PhysRevResearch.4.043113,komelj2025quantum, austrup2025dynamics}.

\begin{acknowledgments}
 We thank T. Tomita and I. Danshita for their useful comments. This work was supported by JSPS KAKENHI Grants No. JP22H05268 (M.K.), No.~JP25K00215 (M.K.), and JST ASPIRE No.~JPMJAP24C2 (M.K.).
 
\end{acknowledgments}
\appendix
\clearpage
\begin{widetext}
\section{Relation between site-dependent spin rotation and DM interaction}\label{sec:appendix_rotation}
This Appendix discusses the relation between the site-dependent spin rotation and DM interaction. We consider the following Hamiltonians:
\begin{align}
    \hat{H}_{XY}=J\sum_{\langle j,k \rangle}(\hat{S}^{x}_{\bm{R}_j}\hat{S}^{x}_{\bm{R}_k}+\hat{S}^{y}_{\bm{R}_j}\hat{S}^{y}_{\bm{R}_k}),\\
    \hat{H}_{YZ}=J\sum_{\langle j,k \rangle}(\hat{S}^{y}_{\bm{R}_j}\hat{S}^{y}_{\bm{R}_k}+\hat{S}^{z}_{\bm{R}_j}\hat{S}^{z}_{\bm{R}_k}),\\
    \hat{H}_{ZX}=J\sum_{\langle j,k \rangle}(\hat{S}^{z}_{\bm{R}_j}\hat{S}^{z}_{\bm{R}_k}+\hat{S}^{x}_{\bm{R}_j}\hat{S}^{x}_{\bm{R}_k}),
\end{align}
where we consider only nearest-neighbor interaction. 
We employ the following spin rotation operator~\cite{PhysRevA.108.053318, kunimi2024}:
\begin{align}
\hat{U}_{{\rm rot},\alpha}\equiv \exp\left(-i\sum_j\hat{S}_{\bm{R}_j}^{\alpha}\phi_j\right),\label{eq:definition_of_rotatinf_operator}
\end{align}
where $\phi_j$ represents the rotation angle at the $j$th site.
By direct calculations, we can obtain the following relations:
\begin{align}
    \hat{U}^{\dagger}_{\rm{rot},z}\hat{H}_{XY}\hat{U}_{\rm{rot},z}=J\sum_{\langle j,k \rangle}\qty[\cos(\phi_j-\phi_k)(\hat{S}^x_{\bm{R}_j}\hat{S}^x_{\bm{R}_k}+\hat{S}^y_{\bm{R}_j}\hat{S}^y_{\bm{R}_k})+\sin(\phi_j-\phi_k)(\hat{S}^x_{\bm{R}_j}\hat{S}^y_{\bm{R}_k}-\hat{S}^y_{\bm{R}_j}\hat{S}^x_{\bm{R}_k})],\\
    \hat{U}^{\dagger}_{\rm{rot},x}\hat{H}_{YZ}\hat{U}_{\rm{rot},x}=J\sum_{\langle j,k \rangle}\qty[\cos(\phi_j-\phi_k)(\hat{S}^y_{\bm{R}_j}\hat{S}^y_{\bm{R}_k}+\hat{S}^z_{\bm{R}_j}\hat{S}^z_{\bm{R}_k})+\sin(\phi_j-\phi_k)(\hat{S}^y_{\bm{R}_j}\hat{S}^z_{\bm{R}_k}-\hat{S}^z_{\bm{R}_j}\hat{S}^y_{\bm{R}_k})],\\
    \hat{U}^{\dagger}_{\rm{rot},y}\hat{H}_{ZX}\hat{U}_{\rm{rot},y}=J\sum_{\langle j,k \rangle}\qty[\cos(\phi_j-\phi_k)(\hat{S}^z_{\bm{R}_j}\hat{S}^z_{\bm{R}_k}+\hat{S}^x_{\bm{R}_j}\hat{S}^x_{\bm{R}_k})+\sin(\phi_j-\phi_k)(\hat{S}^z_{\bm{R}_j}\hat{S}^x_{\bm{R}_k}-\hat{S}^x_{\bm{R}_j}\hat{S}^z_{\bm{R}_k})].
\end{align}
From the above results, by taking $\phi_j-\phi_k=\pi/2$, the DM interaction for each component of the DM vector is obtained.

\section{Derivation of the nearest-neighbor part of the effective Hamiltonian}\label{sec:appendix_derivation_Bloch_and_Neel}
This Appendix discusses the engineering of the Hamiltonian with the Heisenberg and Bloch-type DM interaction in detail. The Hamiltonian with N\'{e}el-type DM interaction can be obtained by a similar calculation.
The target Hamiltonian is
\begin{align}
    \hat{H}^{\mathrm{B}}_{\mathrm{target}}=J\sum_{\langle i,j \rangle}\hat{\bm{S}}_{\bm{R}_{i}}\cdot\hat{\bm{S}}_{\bm{R}_{j}}+\sum_{i}\Big[D_x(\hat{\bm{S}}_{\bm{R}_{i}}\times\hat{\bm{S}}_{\bm{R}_{i}+a\bm{e}_x}){}_x+D_y(\hat{\bm{S}}_{\bm{R}_{i}}\times\hat{\bm{S}}_{\bm{R}_{i}+a\bm{e}_y}){}_y\Big].
\end{align}

The Hamiltonian in the laboratory frame is given by
\begin{align}
    \hat{H}(t)&=\hat{H}_{XY}+\hat{H}_{\rm drive}(t),\label{eq:detinition_of_H(t)_lab_frame_app}\\
    \hat{H}_{\rm{drive}}(t)&=\hat{H}_{\rm{G}}(t)+\hat{H}_{\rm{L}}(t)\notag \\
    &=\hbar\Omega(t)\sum_i[\cos\phi(t)\hat{S}^x_{\bm{R}_i}+\sin\phi(t)\hat{S}^y_{\bm{R}_i}]+\hbar\sum_i\Delta_i(t)\hat{S}^z_{\bm{R}_i},
\end{align}
where $\hat{H}_{\rm{G}}(t)$ is the global pulse term and $\hat{H}_{\rm{L}}(t)$ is the local pulse term. Here, the global pulse is chosen as in the case of the {\it XYZ} Hamiltonian~\cite{scholl2022microwave,geier2021floquet}, and thus the sequence becomes $(\hat{S}^x,-\hat{S}^y,\hat{S}^y,-\hat{S}^x)$ of four $\pi/2$ microwave pulses. The unitary operator associated with the global pulse Hamiltonian $\hat{H}_{\rm{G}}(t)$ is given by
\begin{align}
    \hat{U}_{\rm {G}}(t)= \mathcal{T}\exp\qty[-\frac{i}{\hbar}\int_0^t dt' \hat{H}_{\mathrm G}(t')],
\end{align}
\hk{
where $\mathcal{T}$ represents the time-ordering operator. The direct calculation yields 
\begin{align}
\hat{U}_{\rm{G}}(t)&=
\begin{cases}
\hat{U}_1\equiv \hat{1},&0\le t<\tau_1,\\
\hat{U}_2\equiv e^{-i\hat{S}_{\rm tot}^x\pi/2}\hat{U}_1=e^{-i\hat{S}_{\rm tot}^x\pi/2},& \tau_1\le t<\tau_1+\cdots+\tau_4,\\
\hat{U}_3\equiv e^{+i\hat{S}_{\rm tot}^y\pi/2}\hat{U}_2=e^{+i\hat{S}_{\rm tot}^y\pi/2}e^{-i\hat{S}_{\rm tot}^x\pi/2},& \tau_1+\cdots+\tau_4\le t<\tau_1+\cdots+\tau_{10},\\
\hat{U}_4\equiv e^{-i\hat{S}_{\rm tot}^y\pi/2}\hat{U}_3=e^{-i\hat{S}_{\rm rot}^x\pi/2},& \tau_1+\cdots+\tau_{10}\le t<\tau_1+\cdots+\tau_{13},\\
\hat{U}_5\equiv e^{+i\hat{S}_{\rm tot}^x\pi/2}\hat{U}_4=\hat{1},& \tau_1+\cdots+\tau_{13}\le t<\tau_1+\cdots+\tau_{14}=T.
\end{cases}
\end{align}
}
The Hamiltonian in the first rotating frame is then given by
\begin{align}
    \hat{H}_{\mathrm 1}(t)\equiv\hat{U}^{\dagger}_{\mathrm G}(t)[\hat{H}_{XY}+\hat{H}_{\mathrm L}(t)]\hat{U}_{\mathrm G}(t)\equiv\hat{H}_{XY1}(t)+\hat{H}_{\mathrm L1}(t),
\end{align}
where
\hk{
\begin{align}
    \hat{H}_{XY1}(t)&=
        \begin{cases}
            \hat{U}_1^{\dagger}\hat{H}_0\hat{U}_1=\hat{H}_{XY},& 0\le t<\tau_1,\\
            \hat{U}_2^{\dagger}\hat{H}_0\hat{U}_2=e^{+i\hat{S}_{\rm tot}^x\pi/2}\hat{H}_0e^{-i\hat{S}_{\rm tot}^x\pi/2}=\hat{H}_{ZX},& \tau_1\le t<\tau_1+\cdots+\tau_4,\\
            \hat{U}_3^{\dagger}\hat{H}_0\hat{U}_3=e^{+i\hat{S}_{\rm tot}^x\pi/2}e^{-i\hat{S}_{\rm tot}^y\pi/2}\hat{H}_0e^{+i\hat{S}_{\rm tot}^y\pi/2}e^{-i\hat{S}_{\rm tot}^x\pi/2}=\hat{H}_{YZ},& \tau_1+\cdots+\tau_4\le t<\tau_1+\cdots+\tau_{10},\\
            \hat{U}^{\dagger}_4\hat{H}_0\hat{U}_4=e^{+i\hat{S}_{\rm rot}^x\pi/2}\hat{H}_0e^{-i\hat{S}_{\rm rot}^x\pi/2}=\hat{H}_{ZX},& \tau_1+\cdots+\tau_{10}\le t<\tau_1+\cdots+\tau_{13},\\
            \hat{U}_5^{\dagger}\hat{H}_0\hat{U}_5=\hat{H}_{XY},& \tau_1+\cdots+\tau_{13}\le t<\tau_1+\cdots+\tau_{14}=T.
        \end{cases}\\
    \hat{H}_{\mathrm L1}(t)&=\hat{U}^{\dagger}_{\mathrm G}(t)\hat{H}_{\mathrm L}(t)\hat{U}_{\mathrm G}(t).
\end{align}
}
Next, we consider the second rotating frame. In this frame, the Hamiltonian is given by
\begin{align}
  \hat{H}_{\mathrm 2}(t)&=\hat{U}^{\dagger}_{\mathrm L}(t)\hat{H}_{XY1}(t)\hat{U}_{\mathrm L}(t),\label{eq:definition_of_Hamiltonian_in_second_rotating_frame}\\
  \hat{U}_{\rm L}(t)&\equiv\mathcal{T}\exp\left[-\frac{i}{\hbar}\int^t_0dt'\hat{H}_{\rm L1}(t')\right].\label{eq:definition_of_time_evolution_operator_local_pulses}
\end{align}
Since our pulse sequence is piecewise constant in time, $\hat{U}_{\rm L}(t)$ can be written as
\hk{\begin{align}
\hat{U}_{\rm L}(t)&=\hat{\mathcal{V}}_{k},\quad \text{for } \sum_{i=1}^k \tau_i\le t<\sum_{i=1}^{k+1}\tau_{i},\label{eq:rewrite_U_L}\\
\hat{\mathcal{V}}_{k}&\equiv\hat{V}_{k}\hat{V}_{k-1}\cdots\hat{V}_{1}\hat{V}_{0},\label{eq:definition_of_accumurate_V}
\end{align}}
where $\hat{V}_{k}$ is the time-evolution operator at time $\tau_k$ in the first rotating frame.
The effective Hamiltonian is given by
\begin{align}
\hat{H}_{\rm F}^{(0)}&\equiv \frac{1}{T}\int^T_0dt\hat{H}_2(t)\notag \\
&=\frac{1}{T}\sum_{k=1}^{14}\hat{\mathcal{V}}^{\dagger}_{k-1}\hat{\mathcal{H}}_k\hat{\mathcal{V}}_{k-1}\tau_k,\label{eq:general_form_of_Floquet_Hamiltonian_app}
\end{align}
where $\hat{\mathcal{H}}_k$ represents $\hat{H}_{XY1}(t)$ during the time interval $\tau_k$. From these results, we choose $\hat{\mathcal{V}}_j$ to create the target Hamiltonian. The results are given by
\hk{
\begin{align}
\hat{U}_{\mathrm L}(t)&=\begin{cases}
     \hat{\mathcal{V}}_{0}=\hat{1}, & 0\leq t <\tau_1,\\   
     \hat{\mathcal{V}}_{1}=\hat{1}, & \tau_1\leq t <\tau_1+\tau_2,\\
     \hat{\mathcal{V}}_{2}=\exp\Big\{-i\sum_{j}\hat{S}^{y}_{\bm{R}_{j}}[2\pi-(n_j+m_j)\pi/2]\Big\}, & \tau_1+\tau_2 \leq t <\tau_1+\cdots+\tau_3,\\
     \hat{\mathcal{V}}_{3}=\hat{1}, & \tau_1+\cdots+\tau_3\leq t <\tau_1+\cdots+\tau_4,\\
     \hat{\mathcal{V}}_{4}=\hat{1}, & \tau_1+\cdots+\tau_4 \leq t <\tau_1+\cdots+\tau_5,\\
     \hat{\mathcal{V}}_{5}=\exp\Big\{-i\sum_{j}\hat{S}^{x}_{\bm{R}_{j}}[2\pi-(n_j+m_j)\pi/2]\Big\} & \tau_1+\cdots+\tau_5 \leq t <\tau_1+\cdots+\tau_6,\\
     \hat{\mathcal{V}}_{6}=\hat{1}, & \tau_1+\cdots+\tau_6 \leq t <\tau_1+\cdots+\tau_7,\\
     \hat{\mathcal{V}}_{7}=\hat{1}, & \tau_1+\cdots+\tau_7 \leq t <\tau_1+\cdots+\tau_8,\\
     \hat{\mathcal{V}}_{8}=\exp\Big\{-i\sum_{j}\hat{S}^{x}_{\bm{R}_{j}}[2\pi+(m_j-n_j)\pi/2]\Big\}, & \tau_1+\cdots+\tau_8 \leq t <\tau_1+\cdots+\tau_9,\\
     \hat{\mathcal{V}}_{9}=\hat{1}, & \tau_1+\cdots+\tau_9 \leq t <\tau_1+\cdots+\tau_{10},\\
     \hat{\mathcal{V}}_{10}=\hat{1}, & \tau_1+\cdots+\tau_{10} \leq t <\tau_1+\cdots+\tau_{11},\\
     \hat{\mathcal{V}}_{11}=\exp\Big\{-i\sum_{j}\hat{S}^{y}_{\bm{R}_{j}}[2\pi+(n_j-m_j)\pi/2]\Big\}, & \tau_1+\cdots+\tau_{11} \leq t <\tau_1+\cdots+\tau_{12},\\
     \hat{\mathcal{V}}_{12}=\hat{1}, & \tau_1+\cdots+\tau_{12} \leq t <\tau_1+\cdots+\tau_{13},\\
     \hat{\mathcal{V}}_{13}=\hat{1}, & \tau_1+\cdots+\tau_{13} \leq t <T.
  \end{cases}
\end{align}
}
Using Eq.~(\ref{eq:definition_of_accumurate_V}), we obtain the expressions for $\hat{V}_j$:
\begin{align}
    \hat{V}_{0}&=\hat{1},\\
    \hat{V}_{1}&=\hat{1},\\
    \hat{V}_{2}&=\exp\Big\{-i\sum_{j}\hat{S}^{y}_{\bm{R}_{j}}[2\pi-(n_j+m_j)\pi/2]\Big\},\\
    \hat{V}_{3}&=\exp\Big\{+i\sum_{j}\hat{S}^{y}_{\bm{R}_{j}}[2\pi-(n_j+m_j)\pi/2]\Big\},\\
    \hat{V}_{4}&=\hat{1},\\
    \hat{V}_{5}&=\exp\Big\{-i\sum_{j}\hat{S}^{x}_{\bm{R}_{j}}[2\pi-(n_j+m_j)\pi/2]\Big\},\\
    \hat{V}_{6}&=\exp\Big\{+i\sum_{j}\hat{S}^{x}_{\bm{R}_{j}}[2\pi-(n_j+m_j)\pi/2]\Big\},\\
    \hat{V}_{7}&=\hat{1},\\
    \hat{V}_{8}&=\exp\Big\{-i\sum_{j}\hat{S}^{x}_{\bm{R}_{j}}[2\pi+(m_j-n_j)\pi/2]\Big\},\\
    \hat{V}_{9}&=\exp\Big\{+i\sum_{j}\hat{S}^{x}_{\bm{R}_{j}}[2\pi+(m_j-n_j)\pi/2]\Big\},\\
    \hat{V}_{10}&=\hat{1},\\
    \hat{V}_{11}&=\exp\Big\{-i\sum_{j}\hat{S}^{y}_{\bm{R}_{j}}[2\pi+(n_j-m_j)\pi/2]\Big\},\\
    \hat{V}_{12}&=\exp\Big\{+i\sum_{j}\hat{S}^{y}_{\bm{R}_{j}}[2\pi+(n_j-m_j)\pi/2]\Big\},\\
    \hat{V}_{13}&=\hat{1}.
\end{align}
The effective Hamiltonian becomes 
\begin{align}
    \notag \hat{H}^{(0)}_{\rm{F}}&=\frac{1}{T}\sum_{k=1}^{14}\hat{\mathcal{V}}^{\dagger}_{k-1}\hat{\mathcal{H}}_k\hat{\mathcal{V}}_{k-1}\tau_k\\
    &=\alpha_{XY}\hat{H}_{XY}+\alpha_{YZ}\hat{H}_{YZ}+\alpha_{XZ}\hat{H}_{ZX}+\sum_{i}\frac{J}{T}\Big[(\tau_6+\tau_9)(\hat{\bm{S}}_{\bm{R}_{i}}\times\hat{\bm{S}}_{\bm{R}_{i}+a\bm{e}_x}){}_x\\
    \notag&+(\tau_
    3-\tau_{12})(\hat{\bm{S}}_{\bm{R}_{i}}\times\hat{\bm{S}}_{\bm{R}_{i}+a\bm{e}_x}){}_y+(\tau_6-\tau_9)(\hat{\bm{S}}_{\bm{R}_{i}}\times\hat{\bm{S}}_{\bm{R}_{i}+a\bm{e}_y}){}_x+(\tau_3+\tau_{12})(\hat{\bm{S}}_{\bm{R}_{i}}\times\hat{\bm{S}}_{\bm{R}_{i}+a\bm{e}_y}){}_y \Big].
\end{align}
To implement the target Hamiltonian, we set the propagation times as $\tau_1+\tau_{14}=\tau_2+\tau_4+\tau_{11}+\tau_{13}=\tau_5+\tau_7+\tau_8+\tau_{10}=\tau$, $\tau_6=\tau_9=\tau_x$, and $\tau_3=\tau_{12}=\tau_y$. The effective Hamiltonian then reduces to the target Hamiltonian:
\begin{align}
    \hat{H}^{(0)}_{\rm{F}}=J_{\rm{F}}\sum_{\langle i,j \rangle}\hat{\bm{S}}_{\bm{R}_{i}}\cdot\hat{\bm{S}}_{\bm{R}_{j}}+\sum_i\qty[D_{x,\rm{F}}(\hat{\bm{S}}_{\bm{R}_{i}}\times\hat{\bm{S}}_{\bm{R}_{i}+a\bm{e}_x}){}_x+D_{y,\rm{F}}(\hat{\bm{S}}_{\bm{R}_{i}}\times\hat{\bm{S}}_{\bm{R}_{i}+a\bm{e}_y}){}_y].
\end{align} 

Finally, we consider the transformation of the time-evolution operator in the rotating frame to the laboratory frame, which is required for the experiments. We consider the time-evolution operator given by
\hk{\begin{align}
\hat{U}(t)&=\mathcal{T}\exp\left[-\frac{i}{\hbar}\int^t_0dt'\hat{H}_{\rm drive}(t')\right]\notag \\
&=\hat{\mathcal{P}}_k,\quad \text{for }\sum_{i=1}^{k}\tau_i\le t<\sum_{i=1}^{k+1}\tau_{i},\label{eq:time_evolution_operator_lab_frame}\\
\hat{\mathcal{P}}_k&\equiv\hat{P}_k\hat{P}_{k-1}\cdots\hat{P}_1\hat{P}_0,\label{eq:definition_of_accumurate_P}
\end{align}}
where we used the fact that the pulse sequence is a piecewise function of time. Using $\hat{H}_{\rm drive}(t)=\hat{H}_{\rm G}(t)+\hat{U}_{\rm G}(t)\hat{H}_{\rm L1}(t)\hat{U}^{\dagger}_{\rm G}(t)$, we obtain the time-evolution operator in the laboratory frame as
\hk{\begin{align}
    \hat{U}(t)
      &=\begin{cases}
        \hat{\mathcal{P}}_{0}=\hat{1} & 0\leq t <\tau_1,\\
        \hat{\mathcal{P}}_{1}=\hat{P}_{1} & \tau_1\leq t <\tau_1+\tau_2,\\
        \hat{\mathcal{P}}_{2}=\hat{P}_{2}\hat{P}_{1} & \tau_1+\tau_2 \leq t <\tau_1+\cdots+\tau_3,\\
        \hat{\mathcal{P}}_{3}=\hat{P}_{1} & \tau_1+\cdots+\tau_3\leq t <\tau_1+\cdots+\tau_4,\\
        \hat{\mathcal{P}}_{4}=\hat{P}_{4}\hat{P}_{1} & \tau_1+\cdots+\tau_4 \leq t <\tau_1+\cdots+\tau_5,\\
        \hat{\mathcal{P}}_{5}=\hat{P}_{5}\hat{P}_{4}\hat{P}_{1} & \tau_1+\cdots+\tau_5 \leq t <\tau_1+\cdots+\tau_6,\\
        \hat{\mathcal{P}}_{6}=\hat{P}_{4}\hat{P}_{1} & \tau_1+\cdots+\tau_6 \leq t <\tau_1+\cdots+\tau_7,\\
        \hat{\mathcal{P}}_{7}=\hat{P}_{4}\hat{P}_{1} &  \tau_1+\cdots+\tau_7\leq t <\tau_1+\cdots+\tau_8,\\
        \hat{\mathcal{P}}_{8}=\hat{P}_{8}\hat{P}_{4}\hat{P}_{1} & \tau_1+\cdots+\tau_8 \leq t <\tau_1+\cdots+\tau_9,\\
        \hat{\mathcal{P}}_{9}=\hat{P}_{4}\hat{P}_{1} & \tau_1+\cdots+\tau_9 \leq t <\tau_1+\cdots+\tau_{10},\\
        \hat{\mathcal{P}}_{10}=\hat{P}_{1} & \tau_1+\cdots+\tau_{10} \leq t <\tau_1+\cdots+\tau_{11},\\
        \hat{\mathcal{P}}_{11}=\hat{P}_{11}\hat{P}_{1} & \tau_1+\cdots+\tau_{11} \leq t <\tau_1+\cdots+\tau_{12},\\
        \hat{\mathcal{P}}_{12}=\hat{P}_{1} & \tau_1+\cdots+\tau_{12} \leq t <\tau_1+\cdots+\tau_{13},\\
        \hat{\mathcal{P}}_{13}=\hat{1} & \tau_1+\cdots+\tau_{13} \leq t <T.
  \end{cases}
\end{align}
}
The operators $\hat{P}_k$ are defined as follows:
\begin{align}
    &\hat{P}_{1}=e^{-i\hat{S}^x_{\rm{tot}}\pi/2},\\
    &\hat{P}_{2}=e^{-i\hat{S}^x_{\rm{tot}}\pi/2}\hat{V}_{2}e^{i\hat{S}^x_{\rm{tot}}\pi/2}=\exp\Big\{-i\sum_{j}\hat{S}^{z}_{\bm{R}_{j}}[2\pi-(n_j+m_j)\pi/2]\Big\},\\
    &\hat{P}_{3}=e^{-i\hat{S}^x_{\rm{tot}}\pi/2}\hat{V}_{3}e^{i\hat{S}^x_{\rm{tot}}\pi/2}=\hat{P}^{\dagger}_{2},\\
    &\hat{P}_{4}=e^{i\hat{S}^y_{\rm{tot}}\pi/2},\\
    &\hat{P}_{5}=e^{i\hat{S}^y_{\rm{tot}}\pi/2}e^{-i\hat{S}^x_{\rm{tot}}\pi/2}\hat{V}_{5}e^{i\hat{S}^x_{\rm{tot}}\pi/2}e^{-i\hat{S}^y_{\rm{tot}}\pi/2}=\hat{P}_{2},\\
    &\hat{P}_{6}=e^{i\hat{S}^y_{\rm{tot}}\pi/2}e^{-i\hat{S}^x_{\rm{tot}}\pi/2}\hat{V}_{6}e^{i\hat{S}^x_{\rm{tot}}\pi/2}e^{-i\hat{S}^y_{\rm{tot}}\pi/2}=\hat{P}_{3},\\
    &\hat{P}_{7}=\hat{1},\\
    &\hat{P}_{8}=e^{i\hat{S}^y_{\rm{tot}}\pi/2}e^{-i\hat{S}^x_{\rm{tot}}\pi/2}\hat{V}_{8}e^{i\hat{S}^x_{\rm{tot}}\pi/2}e^{-i\hat{S}^y_{\rm{tot}}\pi/2}=\exp\Big\{-i\sum_{j}\hat{S}^{z}_{\bm{R}_{j}}[2\pi+(m_j-n_j)\pi/2]\Big\},\\
    &\hat{P}_{9}=e^{i\hat{S}^y_{\rm{tot}}\pi/2}e^{-i\hat{S}^x_{\rm{tot}}\pi/2}\hat{V}_{9}e^{i\hat{S}^x_{\rm{tot}}\pi/2}e^{-i\hat{S}^y_{\rm{tot}}\pi/2}=\hat{P}^{\dagger}_{8},\\
    &\hat{P}_{10}=\hat{P}^{\dagger}_{4},\\
    &\hat{P}_{11}=e^{-i\hat{S}^x_{\rm{tot}}\pi/2}\hat{V}_{11}e^{i\hat{S}^x_{\rm{tot}}\pi/2}=\exp\Big\{-i\sum_{j}\hat{S}^{z}_{\bm{R}_{j}}[2\pi+(n_j-m_j)\pi/2]\Big\},\\
    &\hat{P}_{12}=e^{-i\hat{S}^x_{\rm{tot}}\pi/2}\hat{V}_{12}e^{i\hat{S}^x_{\rm{tot}}\pi/2}=\hat{P}^{\dagger}_{11},\\
    &\hat{P}_{13}=\hat{P}^{\dagger}_{1}.
\end{align}

\section{Derivation of the long-range interaction terms}\label{sec:long-range}
In the main text, we assumed the nearest-neighbor interaction for simplicity in obtaining the effective Hamiltonian. However, the interaction between Rydberg atoms is, in fact, a long-range interaction because of the resonant dipole-dipole interaction. Therefore, in this Appendix, we consider long-range interactions beyond the nearest neighbor.

\subsection{Next-nearest-neighbor interaction}
First, we restrict our consideration to next-nearest neighbor interactions.
The starting Hamiltonian is given by 
\begin{align}
    \hat{H}^{\rm{NNN}}_{XY}=\sum_{\sigma\in\{0,1\}}\sum_i J_{\bm{R}_i, \bm{R}^{\rm{NNN}}_{i,\sigma}}(\hat{S}^x_{\bm{R}_i}\hat{S}^x_{\bm{R}^{\rm{NNN}}_{i,\sigma}}+\hat{S}^y_{\bm{R}_i}\hat{S}^y_{\bm{R}^{\rm{NNN}}_{i,\sigma}}),
\end{align}
where $\bm{R}^{\rm{NNN}}_{i,\sigma}=a\qty([n_i+1]\bm{e}_x+[m_i+(-1)^{\sigma}]\bm{e}_y)$.
Next, we apply the same sequence as described in Sec.~\ref{sec:results} of the main text. The Hamiltonian in the first rotating frame is written as
\begin{align}
    \hat{H}^{\rm{NNN}}_{\mathrm 1}(t)\equiv\hat{U}^{\dagger}_{\mathrm G}(t)[\hat{H}^{\rm{NNN}}_{XY}+\hat{H}_{\mathrm L}(t)]\hat{U}_{\mathrm G}(t)\equiv\hat{H}^{\rm{NNN}}_{XY1}(t)+\hat{H}_{\mathrm L1}(t),
\end{align}
where
\hk{\begin{align}
    \hat{H}^{\rm{NNN}}_{XY1}(t)&=
        \begin{cases}
            \hat{H}^{\rm{NNN}}_{XY}, & 0\leq t <\tau_1,\\
            \hat{H}^{\rm{NNN}}_{\mathrm ZX}, & \tau_1 \leq t <\tau_1+\cdots+\tau_4,\\
            \hat{H}^{\rm{NNN}}_{\mathrm YZ}, & \tau_1+\cdots+\tau_4\leq t <\tau_1+\cdots+\tau_{10},\\
            \hat{H}^{\rm{NNN}}_{\mathrm ZX}, & \tau_1+\cdots+\tau_{10} \leq t <\tau_1+\cdots+\tau_{13},\\
            \hat{H}^{\rm{NNN}}_{XY}, & \tau_1+\cdots+\tau_{13} \leq t <T,
        \end{cases}
\end{align}}
with
\begin{align}
    \hat{H}^{\rm{NNN}}_{XY}=\sum_{\sigma\in\{0,1\}}\sum_i J_{\bm{R}_i, \bm{R}^{\rm{NNN}}_{i,\sigma}}(\hat{S}^x_{\bm{R}_i}\hat{S}^x_{\bm{R}^{\rm{NNN}}_{i,\sigma}}+\hat{S}^y_{\bm{R}_i}\hat{S}^y_{\bm{R}^{\rm{NNN}}_{i,\sigma}}),\\
    \hat{H}^{\rm{NNN}}_{YZ}=\sum_{\sigma\in\{0,1\}}\sum_i J_{\bm{R}_i, \bm{R}^{\rm{NNN}}_{i,\sigma}}(\hat{S}^y_{\bm{R}_i}\hat{S}^y_{\bm{R}^{\rm{NNN}}_{i,\sigma}}+\hat{S}^z_{\bm{R}_i}\hat{S}^z_{\bm{R}^{\rm{NNN}}_{i,\sigma}}),\\
    \hat{H}^{\rm{NNN}}_{ZX}=\sum_{\sigma\in\{0,1\}}\sum_i J_{\bm{R}_i, \bm{R}^{\rm{NNN}}_{i,\sigma}}(\hat{S}^z_{\bm{R}_i}\hat{S}^z_{\bm{R}^{\rm{NNN}}_{i,\sigma}}+\hat{S}^x_{\bm{R}_i}\hat{S}^x_{\bm{R}^{\rm{NNN}}_{i,\sigma}}).
\end{align}
The Hamiltonian in the second rotating frame becomes
\hk{\begin{align}
    \hat{H}^{\rm{NNN}}_{\mathrm 2}(t)&\equiv\hat{U}^{\dagger}_{\mathrm L}(t)\hat{H}^{\rm{NNN}}_{XY1}(t)\hat{U}_{\mathrm L}(t)\\
    &=\begin{cases}
     \hat{H}^{\rm{NNN}}_{XY}, & 0\leq t <\tau_1,\\
     \hat{H}^{\rm{NNN}}_{ZX}, & \tau_1 \leq t <\tau_1+\tau_2,\\
     \sum_{\sigma\in\{0,1\}}\sum_i (-1)^{\sigma+1}J_{\bm{R}_i, \bm{R}^{\rm{NNN}}_{i,\sigma}}(\hat{S}^z_{\bm{R}_i}\hat{S}^z_{\bm{R}^{\rm{NNN}}_{i,\sigma}}+\hat{S}^x_{\bm{R}_i}\hat{S}^x_{\bm{R}^{\rm{NNN}}_{i,\sigma}}), & \tau_1+\tau_2\leq t <\tau_1+\cdots+\tau_3,\\
     \hat{H}^{\rm{NNN}}_{ZX}, & \tau_1+\cdots+\tau_3 \leq t <\tau_1+\cdots+\tau_4,\\
     \hat{H}^{\rm{NNN}}_{YZ}, & \tau_1+\cdots+\tau_4 \leq t <\tau_1+\cdots+\tau_5,\\
     \sum_{\sigma\in\{0,1\}}\sum_i (-1)^{\sigma+1}J_{\bm{R}_i, \bm{R}^{\rm{NNN}}_{i,\sigma}}(\hat{S}^y_{\bm{R}_i}\hat{S}^y_{\bm{R}^{\rm{NNN}}_{i,\sigma}}+\hat{S}^z_{\bm{R}_i}\hat{S}^z_{\bm{R}^{\rm{NNN}}_{i,\sigma}}), & \tau_1+\cdots+\tau_5 \leq t <\tau_1+\cdots+\tau_6,\\
     \hat{H}^{\rm{NNN}}_{YZ}, & \tau_1+\cdots+\tau_6 \leq t <\tau_1+\cdots+\tau_7,\\
     \hat{H}^{\rm{NNN}}_{YZ}, & \tau_1+\cdots+\tau_7 \leq t <\tau_1+\cdots+\tau_8,\\
     \sum_{\sigma\in\{0,1\}}\sum_i (-1)^{\sigma}J_{\bm{R}_i, \bm{R}^{\rm{NNN}}_{i,\sigma}}(\hat{S}^y_{\bm{R}_i}\hat{S}^y_{\bm{R}^{\rm{NNN}}_{i,\sigma}}+\hat{S}^z_{\bm{R}_i}\hat{S}^z_{\bm{R}^{\rm{NNN}}_{i,\sigma}}), & \tau_1+\cdots+\tau_8 \leq t <\tau_1+\cdots+\tau_9,\\
     \hat{H}^{\rm{NNN}}_{YZ}, & \tau_1+\cdots+\tau_9 \leq t <\tau_1+\cdots+\tau_{10},\\
     \hat{H}^{\rm{NNN}}_{ZX}, & \tau_1+\cdots+\tau_{10} \leq t <\tau_1+\cdots+\tau_{11},\\
     \sum_{\sigma\in\{0,1\}}\sum_i (-1)^{\sigma}J_{\bm{R}_i, \bm{R}^{\rm{NNN}}_{i,\sigma}}(\hat{S}^z_{\bm{R}_i}\hat{S}^z_{\bm{R}^{\rm{NNN}}_{i,\sigma}}+\hat{S}^x_{\bm{R}_i}\hat{S}^x_{\bm{R}^{\rm{NNN}}_{i,\sigma}}), & \tau_1+\cdots+\tau_{11} \leq t <\tau_1+\cdots+\tau_{12},\\
     \hat{H}^{\rm{NNN}}_{ZX}, & \tau_1+\cdots+\tau_{12} \leq t <\tau_1+\cdots+\tau_{13},\\
     \hat{H}^{\rm{NNN}}_{XY}, & \tau_1+\cdots+\tau_{13} \leq t <T.
  \end{cases}
\end{align}}
Here, we set the propagation time as in Sec.~\ref{sec:results} of the main text. The effective Hamiltonian restricted to next-nearest neighbor interactions is given by
\begin{align}
    \hat{H}^{(0),\rm{NNN}}_{\rm{F}}=\sum_{\sigma\in\{0,1\}}\sum_i J^{\rm{NNN}}_{\rm{F}}(\hat{\bm{S}}_{\bm{R}_i}\cdot\hat{\bm{S}}_{\bm{R}^{\rm{NNN}}_{i,\sigma}}),
\end{align}
where $J^{\rm{NNN}}_{\rm{F}}\equiv 2\tau J_{\bm{R}_i, \bm{R}^{\rm{NNN}}_{i,\sigma}}/T$.
Therefore, when considering the next-nearest-neighbor interaction, we find that only the Heisenberg interaction appears. The reason why the DM interaction is absent is that the phase difference between site $\bm{R}_i$ and site $\bm{R}^{\rm{NNN}}_{i,\sigma}$ is always $\pm\pi$ (see Appendix~\ref{sec:appendix_rotation}). 

\subsection{Next-next-nearest-neighbor interaction}
Next, we restrict our consideration to next-next-nearest-pneighbor interactions.
The starting Hamiltonian is given by
\begin{align}
    \hat{H}^{\rm{NNNN}}_{XY}=\sum_{\mu\in\{x,y\}}\sum_i J_{\bm{R}_i, \bm{R}^{\rm{NNNN}}_{i,\mu}}(\hat{S}^x_{\bm{R}_i}\hat{S}^x_{\bm{R}^{\rm{NNNN}}_{i,\mu}}+\hat{S}^y_{\bm{R}_i}\hat{S}^y_{\bm{R}^{\rm{NNNN}}_{i,\mu}}),
\end{align}
where $\bm{R}^{\rm{NNNN}}_{i,\mu}=\bm{R}_{i}+2a\bm{e}_{\mu}$.
Next, we apply the same sequence as described in the main text. The Hamiltonian in the first rotating frame is written as
\begin{align}
    \hat{H}^{\rm{NNNN}}_{\mathrm 1}(t)\equiv\hat{U}^{\dagger}_{\mathrm G}(t)[\hat{H}^{\rm{NNNN}}_{XY}+\hat{H}_{\mathrm L}(t)]\hat{U}_{\mathrm G}(t)\equiv\hat{H}^{\rm{NNNN}}_{XY1}(t)+\hat{H}_{\mathrm L1}(t),
\end{align}
where
\hk{\begin{align}
    \hat{H}^{\rm{NNNN}}_{XY1}(t)&=
        \begin{cases}
            \hat{H}^{\rm{NNNN}}_{XY}, & 0\leq t <\tau_1,\\
            \hat{H}^{\rm{NNNN}}_{ZX}, & \tau_1 \leq t <\tau_1+\cdots+\tau_4,\\
            \hat{H}^{\rm{NNNN}}_{YZ}, & \tau_1+\cdots+\tau_4\leq t <\tau_1+\cdots+\tau_{10},\\
            \hat{H}^{\rm{NNNN}}_{ZX}, & \tau_1+\cdots+\tau_{10} \leq t <\tau_1+\cdots+\tau_{13},\\
            \hat{H}^{\rm{NNNN}}_{XY}, & \tau_1+\cdots+\tau_{13} \leq t <T,
        \end{cases}
\end{align}}
with 
\begin{align}
    \hat{H}^{\rm{NNNN}}_{XY}\equiv\sum_{\mu\in\{x,y\}}\sum_i J_{\bm{R}_i, \bm{R}^{\rm{NNNN}}_{i,\mu}}(\hat{S}^x_{\bm{R}_i}\hat{S}^x_{\bm{R}^{\rm{NNNN}}_{i,\mu}}+\hat{S}^y_{\bm{R}_i}\hat{S}^y_{\bm{R}^{\rm{NNNN}}_{i,\mu}}), \\
    \hat{H}^{\rm{NNNN}}_{YZ}\equiv\sum_{\mu\in\{x,y\}}\sum_i J_{\bm{R}_i, \bm{R}^{\rm{NNNN}}_{i,\mu}}(\hat{S}^y_{\bm{R}_i}\hat{S}^y_{\bm{R}^{\rm{NNNN}}_{i,\mu}}+\hat{S}^z_{\bm{R}_i}\hat{S}^z_{\bm{R}^{\rm{NNNN}}_{i,\mu}}),\\
    \hat{H}^{\rm{NNNN}}_{ZX}\equiv\sum_{\mu\in\{x,y\}}\sum_i J_{\bm{R}_i, \bm{R}^{\rm{NNNN}}_{i,\mu}}(\hat{S}^z_{\bm{R}_i}\hat{S}^z_{\bm{R}^{\rm{NNNN}}_{i,\mu}}+\hat{S}^x_{\bm{R}_i}\hat{S}^x_{\bm{R}^{\rm{NNNN}}_{i,\mu}}).
\end{align}
We then apply the same pulses again. 
The Hamiltonian in the second rotating frame in each time interval is given by
\hk{\begin{align}
    \hat{H}^{\rm{NNNN}}_{\mathrm 2}(t)&=\hat{U}^{\dagger}_{\mathrm L}(t)\hat{H}^{\rm{NNNN}}_{XY1}(t)\hat{U}_{\mathrm L}(t)\\
    &=\begin{cases}
     \hat{H}^{\rm{NNNN}}_{XY}, & 0\leq t <\tau_1,\\
     \hat{H}^{\rm{NNNN}}_{ZX}, & \tau_1 \leq t <\tau_1+\tau_2,\\
     -\hat{H}^{\rm{NNNN}}_{ZX}, & \tau_1+\tau_2\leq t <\tau_1+\cdots+\tau_3,\\
     \hat{H}^{\rm{NNNN}}_{ZX}, & \tau_1+\cdots+\tau_3 \leq t <\tau_1+\cdots+\tau_4,\\
     \hat{H}^{\rm{NNNN}}_{YZ}, & \tau_1+\cdots+\tau_4 \leq t <\tau_1+\cdots+\tau_5,\\
     -\hat{H}^{\rm{NNNN}}_{YZ}, & \tau_1+\cdots+\tau_5 \leq t <\tau_1+\cdots+\tau_6,\\
     \hat{H}^{\rm{NNNN}}_{YZ}, & \tau_1+\cdots+\tau_6 \leq t <\tau_1+\cdots+\tau_7,\\
     \hat{H}^{\rm{NNNN}}_{YZ}, & \tau_1+\cdots+\tau_7 \leq t <\tau_1+\cdots+\tau_8,\\
     -\hat{H}^{\rm{NNNN}}_{YZ}, & \tau_1+\cdots+\tau_8 \leq t <\tau_1+\cdots+\tau_9,\\
     \hat{H}^{\rm{NNNN}}_{YZ}, & \tau_1+\cdots+\tau_9 \leq t <\tau_1+\cdots+\tau_{10},\\
     \hat{H}^{\rm{NNNN}}_{ZX}, & \tau_1+\cdots+\tau_{10} \leq t <\tau_1+\cdots+\tau_{11},\\
     -\hat{H}^{\rm{NNNN}}_{ZX}, & \tau_1+\cdots+\tau_{11} \leq t <\tau_1+\cdots+\tau_{12},\\
     \hat{H}^{\rm{NNNN}}_{ZX}, & \tau_1+\cdots+\tau_{12} \leq t <\tau_1+\cdots+\tau_{13},\\
     \hat{H}^{\rm{NNNN}}_{XY}, & \tau_1+\cdots+\tau_{13} \leq t <T.\\
  \end{cases}
\end{align}}
Here, we set the propagation time as in the main text. The effective Hamiltonian restricted to next-next-nearest neighbor interactions is given by
\begin{align}
    \hat{H}^{(0),\rm{NNNN}}_{\rm{F}}=\sum_{\mu\in\{x,y\}}\sum_i J^{\rm{NNNN}}_{\rm{F}}\qty[(\tau-\tau_y)\hat{S}^x_{\bm{R}_i}\hat{S}^x_{\bm{R}^{\rm{NNNN}}_{i,\mu}}+(\tau-\tau_x)\hat{S}^y_{\bm{R}_i}\hat{S}^y_{\bm{R}^{\rm{NNNN}}_{i,\mu}}+(\tau-\tau_x-\tau_y)\hat{S}^z_{\bm{R}_i}\hat{S}^z_{\bm{R}^{\rm{NNNN}}_{i,\mu}}],
\end{align}
where $J^{\rm{NNNN}}_{\rm{F}}\equiv 2\tau J_{\bm{R}_i, \bm{R}^{\rm{NNNN}}_{i,\mu}}/T$. The reason why the DM interaction is absent is the same as in the case of the next-nearest-neighbor interaction.

\section{Time evolution of spin expectation values}\label{sec:spin expectation values}
In this Appendix, we consider the Hamiltonian with Bloch-type DM interaction given by
\begin{align}
    \notag \hat{H}^{(0)}_{\rm{F,Bloch}}&=J_{\rm{F}}\sum_{\langle i,j\rangle}\hat{\bm{S}}_{\bm{R}_{i}}\cdot\hat{\bm{S}}_{\bm{R}_{j}}
    \notag +\sum_i\left[D_{x,\rm{F}}(\hat{\bm{S}}_{\bm{R}_{i}}\times\hat{\bm{S}}_{\bm{R}_{i}+a\bm{e}_x}){}_x
    +D_{y,\rm{F}}(\hat{\bm{S}}_{\bm{R}_{i}}\times\hat{\bm{S}}_{\bm{R}_{i}+a\bm{e}_y}){}_y\right]\\
    &\equiv \hat{H}_{\rm ex}+\hat{H}_{\rm{DM}x,x}+\hat{H}_{\rm{DM}y,y}.
\end{align}
Now, we consider a symmetry of the Hamiltonian $\hat{H}^{(0)}_{\rm{F,Bloch}}$.
We define the space-inversion operator along the $y$ direction $\hat{I}_y$ and spin-inversion operators $ \hat{C}_{\mu}$ as
\begin{align}
    \hat{I}^{\dagger}_y\hat{S}^{\mu}_{(R_{j,x},R_{j,y})}\hat{I}_y&\equiv\hat{S}^{\mu}_{(R_{j,x},L_y-R_{j,y}+a)},\\
    \hat{C}_{\mu}&\equiv \prod_j(2\hat{S}_{\bm{R}_j}^{\mu}),
\end{align}
where $\mu=x,y,z$ and $L_y$ represents the number of lattice sites along the $y$ direction. For simplicity, we have assumed that $L_y$ is even. From the above operations, we define the unitary operator
\begin{align}
    \hat{U}_{\rm s}&\equiv \hat{I}_y\hat{C}_{x}.
\end{align}
It is straightforward to show that the unitary operator $\hat{U}_{\rm s}$ commutes with $\hat{H}^{(0)}_{\rm{F,Bloch}}$ and that $\hat{U}^{\dagger}_{\rm s}\hat{S}^{y,z}_{\rm tot}\hat{U}_{\rm s}=-\hat{S}^{y,z}_{\rm tot}$.

We now consider the equation of motion for the expectation values of $\hat{S}_{\rm tot}^{y,z}$ with the initial condition $\ket{\psi(t=0)}\equiv\prod_{j=1}^{N}\ket*{+_j}$: 
\begin{align}
    i\hbar \dv{t}\expval{\hat{S}^{y,z}_{\rm tot}(t)}{\psi(t=0)}&=\expval{e^{i\hat{H}^{(0)}_{\rm{F,Bloch}}t/\hbar}\comm{\hat{S}^{y,z}_{\rm tot}}{\hat{H}^{(0)}_{\rm{F,Bloch}}}e^{-i\hat{H}^{(0)}_{\rm{F,Bloch}}t/\hbar}}{\psi(t=0)}.
    \label{eq:expval time evolution}
\end{align}
Since the initial state $\ket{\psi(t=0)}$ is invariant to the space and spin inversion, one has the relation $\hat{U}_{\rm s}\ket{\psi(t=0)}=\ket{\psi(t=0)}$. 
Using this relation and the symmetry properties of $\hat{H}^{(0)}_{\rm{F,Bloch}}$ and $\hat{S}^{y,z}_{\rm tot}$ with respect to $\hat{U}_{\rm s}$ described above, we obtain 
\begin{align}
    \notag i\hbar \dv{t}\expval{\hat{S}^{y,z}_{\rm tot}(t)}{\psi(t=0)}&=\expval{e^{i\hat{H}^{(0)}_{\rm{F,Bloch}}t/\hbar}\hat{U}^{\dagger}_{\rm s}\comm{\hat{S}^{y,z}_{\rm tot}}{\hat{H}^{(0)}_{\rm{F,Bloch}}}\hat{U}_{\rm s}e^{-i\hat{H}^{(0)}_{\rm{F,Bloch}}t/\hbar}}{\psi(t=0)}\\
    &=-\expval{e^{i\hat{H}^{(0)}_{\rm{F,Bloch}}t/\hbar}\comm{\hat{S}^{y,z}_{\rm tot}}{\hat{H}^{(0)}_{\rm{F,Bloch}}}e^{-i\hat{H}^{(0)}_{\rm{F,Bloch}}t/\hbar}}{\psi(t=0)}.
    \label{eq:d9}
\end{align}
Comparing Eq.~\eqref{eq:d9} with \eqref{eq:expval time evolution}, we conclude that
\begin{align}
    i\hbar \dv{t}\expval{\hat{S}^{y,z}_{\rm tot}(t)}{\psi(t=0)}&=0.
\end{align}
Since the expectation values of $\hat{S}^{y,z}_{\rm tot}$ are zero at the initial time $t=0$, they are always zero.

\end{widetext}

\bibliographystyle{apsrev4-2} 
\bibliography{apsskyrmion} 

\end{document}